\documentclass[letterpaper,twocolumn,10pt]{article}
\usepackage{usenix2019_v3}

\usepackage{tikz}
\usepackage{amsmath}
\usepackage{cite}
\usepackage{tabularx}
\usepackage{ragged2e}
\usepackage{algorithm}
\usepackage[noend]{algpseudocode}
\usepackage{subcaption} 
\usepackage{graphicx}
\usepackage{arydshln}
\usepackage{multirow}
\usepackage{booktabs}
\usepackage{enumitem}
\usepackage[utf8]{inputenc}
\usepackage[T1]{fontenc}
\usepackage{xcolor}
\usepackage{minted}
\usepackage[most]{tcolorbox}
\tcbuselibrary{minted,breakable}
\newtcolorbox{findingbox}{
  width=\linewidth,
  leftright skip=3pt,
  colback=gray!10,
  colframe=gray!80,
  boxrule=0.8pt,
  arc=6pt,
  left=6pt,
  right=6pt,
  top=6pt,
  bottom=6pt,
}
\usepackage{makecell}
\usepackage{amssymb}

\begin{document}

\date{}

\title{\Large \bf Reconstruction of Personally Identifiable Information from Proprietary Data in Supervised Fine-Tuned Models}

\author{
{\rm Sae Furukawa}\\
Northeastern University\\
furukawa.s@northeastern.edu
\and
{\rm Alina Oprea}\\
Northeastern University\\
a.oprea@northeastern.edu
} 

\maketitle
\begin{abstract}
  Supervised Finetuning (SFT) has become one of the primary methods for adapting a large language model (LLM) with extensive pre-trained knowledge to domain-specific, instruction-following tasks. SFT datasets, composed of instruction-response pairs, often include user-provided information that may contain sensitive data such as personally identifiable information (PII), raising privacy concerns. This paper studies the problem of \emph{targeted PII reconstruction} from models finetuned on proprietary SFT data, in which an adversary attempts to recover PII associated with a specific identity. We construct multi-turn, user-centric Q\&A datasets in sensitive domains, specifically medical and legal settings, that incorporate PII to enable realistic evaluation of leakage. We then propose COVA, a coverage-aware decoding algorithm for targeted PII reconstruction under prefix-based attacks. Using COVA, we study how the amount of information available about a target user affects the recovery of their PII from SFT models. Across datasets, COVA consistently improves PII reconstruction over baseline decoding methods, and even limited contextual knowledge can substantially increase an adversary’s success. Our findings demonstrate that small, proprietary SFT datasets can induce meaningful privacy leakage through the reconstruction of target-PII associations learned during fine-tuning.

\end{abstract}
\section{Introduction}
Large Language Models (LLMs) have demonstrated strong capabilities across various language tasks~\cite{gpt, gemini, anthropic2024claude3, zero-shot-learners}. These LLMs, such as the GPT~\cite{gpt} and Claude~\cite{anthropic2024claude3} families, are pre-trained on large-scale text corpora and scaled to billions or trillions of parameters, enabling them to acquire broad knowledge. Supervised Fine-Tuning (SFT)~\cite{sft-instruct} is commonly used to adapt these pre-trained models to domain-specific tasks. SFT trains models on instruction–response pairs, allowing them to generate appropriate outputs conditioned on user queries. Compared to pre-training, SFT typically requires significantly smaller datasets and has proven effective in improving performance on specialized tasks~\cite{zero-shot-learners, sft-abilities, sft-instruct}. 

Memorization in LLMs has been identified as a significant privacy concern, and prior work has shown that attackers with black-box access can extract portions of the training data~\cite{carlini-extraction,quantifying-memorization}. Prior work on Personally Identifiable Information
(PII) extraction spans several settings that differ both in how PII enters the model and in what the attacker seeks to recover. One line of work studies PII contained in large public pre-training corpora~\cite{janus-finetuning, few-shot-prompting, analyzing-pii-leakage} while another evaluates models fine-tuned directly on large, PII-rich public corpora using next-token prediction objectives~\cite{pii-extraction-optimized-prompt, hidden-dangers}. Since these corpora are public and widely used, the same PII may already have appeared during pre-training. Therefore, fine-tuning on such data may only facilitate extraction of previously learned PII, making it difficult to isolate privacy leakage introduced by fine-tuning itself. This setting is also less representative of realistic proprietary SFT datasets, which are collected primarily to improve model performance on downstream tasks rather than to explicitly learn PII-rich corpora. We instead study leakage of PII introduced exclusively through proprietary SFT, where PII appears only incidentally in the downstream task.

Beyond how PII enters the model, prior attacks also differ in their extraction objective. \emph{Untargeted attacks} seek arbitrary PII records, demonstrating that models can expose PII at scale, whereas \emph{targeted attacks} seek PII associated with a specified identity. The targeted setting is particularly relevant to user-centric SFT datasets, where Q\&A records collected through crowdsourcing or user interactions may associate an individual's identity with PII and sensitive context, such as medical symptoms or legal circumstances~\cite{human-chatbot-interactions}. However, existing targeted attacks rely on prerequisites such as leaked target--PII pairs or publicly available target context for prompt augmentation~\cite{janus-finetuning, few-shot-prompting}. These prerequisites may be unavailable when the associations exist only in proprietary SFT data, leaving targeted reconstruction from small, task-oriented datasets underexplored. This raises the following question: \textit{to what extent can PII embedded in proprietary SFT datasets be reconstructed from the resulting fine-tuned model?}

To address this gap, we study \emph{targeted PII reconstruction} from instruction-tuned models trained on small, user-centric, proprietary SFT datasets. We consider threat models with varying levels of attacker knowledge, ranging from knowledge of only a target identity to partial contextual information or a redacted version of the original SFT record. Under each setting, the attacker attempts to reconstruct the PII associated with a specific target rather than extract arbitrary PII strings.

We identify two limitations of decoding approaches commonly used in prior extraction attacks~\cite{analyzing-pii-leakage, janus-finetuning, pii-extraction-optimized-prompt, few-shot-prompting}. First, token-level search can suffer from error propagation: locally suboptimal token choices may eliminate plausible complete PII candidates before their sequence-level likelihood can be evaluated. Second, reconstruction scores are influenced by the model's prior distribution over the target PII type. For example, common dates or locations may receive high probability even when they are unrelated to the target identity, causing prior-driven candidates to outrank memorized associations.

We therefore propose COVA, a coverage-based adversarial decoding method for targeted PII reconstruction that outperforms existing decoding methods. COVA first expands the search over plausible complete candidate strings to improve sequence-level coverage and reduce premature pruning caused by token-level decisions. It then re-ranks the reconstructed candidates by estimating the prior distribution over the target PII type and exploiting the likelihood differential between target-conditioned and prior-driven generation. This separates candidates supported by the target-specific SFT context from candidates that are generally likely under the model's pre-trained prior.

Using this framework, we study how varying levels of attacker knowledge about the target affect the reconstruction of target-PII associations. Our results show that even limited contextual knowledge can substantially increase PII leakage risk, while improved sequence-level decoding exposes leakage that is missed by conventional top-$k$ and beam-search strategies used in prior work~\cite{carlini-extraction,analyzing-pii-leakage, janus-finetuning, pii-extraction-optimized-prompt}.

\noindent \textbf{Contributions.} Our main contributions are as follows:
\begin{itemize}[leftmargin=*]
    \item We propose COVA, a coverage-based adversarial decoding method that improves upon existing decoding methods by reconstructing plausible sequence-level candidates and re-ranking them using an estimated PII prior and target-conditioned likelihood differentials.
    \item We construct multi-turn, user-centric SFT datasets with synthetic PII, built from existing single-turn Q\&A datasets in sensitive domains (medical and legal). 
    \item We show that partial attacker knowledge can improve success in targeted PII reconstruction, even when the attacker does not know the complete SFT record or verbatim prefix.
    \item We analyze how reconstruction performance varies with model scale and family, showing that both architecture and size influence PII leakage.
\end{itemize}

\section{Background}
In this section, we describe key aspects of our research, including the SFT framework, the PII extraction objectives, and prefix-based reconstruction attacks against LLMs.

\subsection{Supervised Fine-tuning}
The supervised fine-tuning (SFT) process utilizes datasets composed of instruction-response $\{(i,r)\}$ where $i$ denotes the input instruction and $r$ is the corresponding target response. Concretely, the SFT objective aims to minimize the negative log-likelihood of the response of $r$ given the context $c$ and instruction $i$:
\begin{equation*}
    \text{L}(\theta) = -\log f_{\theta}(r \mid c, i)
\end{equation*}

where $\theta$ represents the model parameters. The context $c$ may be supplied alongside $i$ and can include a system prompt and additional relevant information, such as the output of information retrieval tools. In a multi-turn Q\&A setting, the context refers to the sequence of question-answer pairs from previous turns. SFT enables models to adapt to domain-specific knowledge, including medicine~\cite{clinical-knowledge} and legal advice~\cite{chatlaw}. SFT may update all model parameters or use parameter-efficient methods such as adapters and low-rank adaptation~\cite{adapters,lora}. We focus on full-parameter SFT.

\subsection{PII Extraction Objectives} 
\noindent \textbf{Personally Identifiable Information.} 
Personally identifiable information (PII) refers to data that can be used to identify or distinguish an individual. It is commonly categorized into \emph{direct identifiers}, which uniquely identify an individual, and \emph{indirect identifiers}, which enable re-identification when combined with other information. Examples of direct identifiers include Social Security numbers and passport numbers. In contrast, indirect identifiers include attributes such as date of birth and place of birth, which may not uniquely identify an individual in isolation but can do so when combined. The sensitivity of PII underscores the importance of studying leakage risks, particularly in light of regulatory frameworks such as the General Data Protection Regulation (GDPR)~\cite{gdpr} and the California Consumer Privacy Act (CCPA)~\cite{ccpa}, which mandate strict protections for personal data and impose significant penalties for non-compliance.

We distinguish between untargeted extraction and targeted PII reconstruction. \emph{Untargeted attacks} attempt to recover any PII string memorized by the model, regardless of the individual to whom it belongs. In contrast, \emph{targeted attacks} begin with an identifier for a specific individual and attempts to recover the corresponding PII. We represent a PII association as [x, y] where (x) represents a target identifier, such as a person's name, and (y) is the PII associated with that individual~\cite{janus-finetuning}. Successful targeted reconstruction therefore requires recovering the correct association, rather than merely generating a valid PII string.

\subsection{Prefix-Based Reconstruction Attacks} 

\noindent \textbf{Prefix Construction.}
In PII extraction frameworks, prefix-based attacks provide the model with a sequence ($S_0$) intended to elicit a memorized continuation~\cite{analyzing-pii-leakage,janus-finetuning,pii-extraction-optimized-prompt,few-shot-prompting}. The prefix may be a verbatim portion of a training example or an attacker-constructed query based on partial knowledge of the target. More target-specific context can increase reconstruction success, but it also represents a stronger attacker-knowledge assumption~\cite{carlini-extraction}. Given a prefix $S_0$ and a candidate string $C=(c_1,\dots,c_{|C|})$, its conditional log-likelihood is
\begin{equation}
\ell_{\theta}(C \mid S_0)
=
\sum_{t=1}^{|C|}
\log P_{\theta}(c_t \mid S_0, c_{<t}).
\end{equation}

The reconstructed PII is then selected as the highest-scoring candidate from the candidate set $\mathcal{C}$:
\begin{equation}
\hat{C}
=
\arg\max_{C \in \mathcal{C}}
\ell_{\theta}(C \mid S_0).
\end{equation}

This ranking compares the likelihood that the model assigns to each candidate continuation.
\section{Problem Statement}
In this paper, we study the SFT setting, where an LLM is adapted to multi-turn chatbot conversations from real user interactions, with users disclosing personally identifiable information (PII)~\cite{human-chatbot-interactions}. We investigate whether an adversary can reconstruct the association between a specific individual and PII introduced through the fine-tuning data. We refer to this as the \emph{targeted PII reconstruction problem} and evaluate it under different levels of attacker knowledge.

\subsection{Attack Objectives and Capabilities}
\noindent \textbf{Scenario.}
\label{scenario}
In real-world deployments, users interact with chatbots through multi-turn conversations, where information accumulates across exchanges. During these interactions, users often disclose personally identifiable information (PII), either explicitly or implicitly, as observed in large-scale analyses of conversational data~\cite{human-chatbot-interactions}. In practice, such user-generated conversation data may be collected and used by companies to further fine-tune models for specialized domains, such as healthcare and legal services. We assume that the chatbot system has persistent user memory, capable of retaining and reusing user-specific information across interactions. This assumption is aligned with modern chatbot systems, such as ChatGPT~\cite{gpt} and Google Gemini~\cite{gemini}, which incorporate memory or user-profile features to provide personalized responses. 

We consider a model provider that begins with a pre-trained language model and performs SFT on a comparatively small, task-oriented dataset of multi-turn conversations. The objective of fine-tuning is to improve performance on a useful downstream task. Nevertheless, PII may appear incidentally in conversations and become associated with the surrounding user-specific context. We focus on this leakage setting in the medical and legal domains.

\noindent \textbf{Attacker Goals.}
\label{attacker_goals}
The attacker seeks to recover PII associated with a user present in the fine-tuning dataset. Given a target identifier, such as the individual's name, together with any available auxiliary context, the attacker attempts to recover the corresponding PII. We refer to this objective as \emph{targeted PII reconstruction}. Our focus is on measuring the leakage of associations between users and their sensitive data. Therefore, we do not consider untargeted attacks that aim to extract arbitrary or unassociated PII at scale. In addition, we focus exclusively on leakage arising from the proprietary fine-tuning dataset and do not consider potential information leakage from the pre-training corpora.

\begin{figure}
    \centering
    \includegraphics[width=1.0\linewidth]{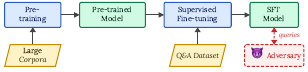}
    \caption{Supervised Finetuning (SFT) Setup}
    \label{fig:sft-pipeline}
\end{figure}

\noindent \textbf{Attacker's Capabilities.}
\label{attacker_capabilities}
We consider an attacker with query-level access to the fine-tuned model, as shown in Figure~\ref{fig:sft-pipeline}. The attacker does not have direct access to model parameters, but can obtain token-level likelihoods, enabling the implementation of custom decoding strategies; this assumption is standard in prior PII extraction work~\cite{analyzing-pii-leakage, pii-extraction-optimized-prompt, carlini-extraction, privacy-ripple-effect}. By default, we assume the attacker does not have access to the pre-trained base model used before finetuning.

\noindent \textbf{Comparison with Prior Works.}
Table~\ref{tab:prior-work-comparison} compares our setting with representative PII extraction studies in terms of attacker access, PII extraction objectives, target models, sources of leaked PII, and key threat-model assumptions. We distinguish \emph{targeted attacks}, which recover PII associated with a specified identity, from \emph{untargeted attacks} that extract arbitrary PII. Among works that study targeted attacks like ours, Chen et al.~\cite{janus-finetuning} require a small set of leaked PII association pairs to construct the malicious fine-tuning data, whereas Cheng et al.~\cite{few-shot-prompting} assume that the target record originates from public corpora, making target-related auxiliary information available for prompt augmentation. Without these prerequisites, the corresponding attacks cannot be directly applied. Another important distinction is that prior work examines PII learned during pre-training or reinforced through fine-tuning on public PII corpora. In contrast, we study targeted reconstruction of PII introduced through proprietary, task-oriented SFT datasets, allowing us to isolate privacy leakage attributable to SFT.

\begin{table*}[t]
\centering
\setlength{\tabcolsep}{1.8pt}
\caption{Comparison with prior PII extraction settings. \textbf{Attacker access} describes the interface exposed by the target model: logit access provides token probabilities, whereas output-only access provides only generated text. \textbf{
Additional assumptions} describes any further prerequisite required for the attack. While prior work examines PII learned during pre-training or reinforced by fine-tuning on large public PII corpora, we study PII leakage introduced through small, proprietary, task-oriented SFT datasets. }
\label{tab:prior-work-comparison}
\footnotesize

\begin{tabular}{
p{0.13\textwidth}
p{0.165\textwidth}
p{0.165\textwidth}
p{0.165\textwidth}
p{0.165\textwidth}
p{0.165\textwidth}}
\hline
& \textbf{Chen et al.~\cite{janus-finetuning}}
& \textbf{Lukas et al.~\cite{analyzing-pii-leakage}}
& \textbf{Keum et al.~\cite{pii-extraction-optimized-prompt}}
& \textbf{Cheng et al.~\cite{few-shot-prompting}}
& \textbf{Our Work} \\
\hline

\textbf{Target model}
& Pre-trained LLM
& Fine-tuned LLM 
& Fine-tuned LLM
& Pre-trained or
& SFT model \\

&
&  
&
& fine-tuned LLM
& \\

\textbf{Attacker access}
& Fine-tuning API
& Logit access
& Logit access
& Output access
& Logit access \\

\textbf{PII extraction type}
& Targeted / untargeted
& Untargeted
& Untargeted
& Targeted / untargeted
& Targeted \\

\textbf{PII source}
& Pre-training data
& Unstructured PII corpora
& Unstructured PII corpora
& Pre-training data
& Proprietary SFT data \\

\textbf{PII types}
& Email, location,
& Email location,
& Email, phone number,
& Email, phone number,
& Email, city, \\

& SSN
& name, phone number
& name
& name
& date of birth \\

\textbf{Additional}
& Small set of leaked PII 
& ---
& A surrogate LM
& Public target context 
& --- \\
\textbf{assumptions}
& association pairs
& 
&
& for prompt augmentation
& \\

\hline
\end{tabular}
\end{table*}

\subsection{Attacker Knowledge}
\label{attacker_knowledge}
The central question we investigate is how varying levels of attacker knowledge affect the ability to infer associations between users and their PII. Accordingly, we consider multiple levels of attacker knowledge. The strongest attacker knowledge assumes full access to the fine-tuning data with PII redacted, while under the weakest assumption the attacker knows only the target identity. We also consider intermediate settings, where the attacker has partial knowledge, such as contextual information (e.g., symptoms or legal dispute), user background attributes (e.g., demographics), or the general format of the dataset. A detailed breakdown of these attacker knowledge settings is provided later in Section~\ref{attacker_knowledge_prefix}.
\section{Attack Methodology}In this section, we describe our methodology for targeted PII reconstruction attacks. Extraction effectiveness can be improved along two conceptually distinct dimensions. \emph{Prompt-side improvements} modify the prefix or prepend additional information to better elicit the target PII, whereas \emph{decoding-side improvements} hold the prefix fixed and change how the model's continuation distribution is explored and ranked. Our attack design focuses primarily on the decoding side: we seek to recover PII associations more effectively given a fixed prefix and attacker knowledge. 

We first present an overview of the attack pipeline. We then describe the attacker-knowledge settings and prefix-construction strategies. Finally, we explain the limitations of standard decoding methods for exact PII reconstruction and introduce COVA, our decoding strategy for addressing these limitations.

\subsection{Attack Pipeline}
\begin{figure}
\centering
\includegraphics[width=\linewidth]{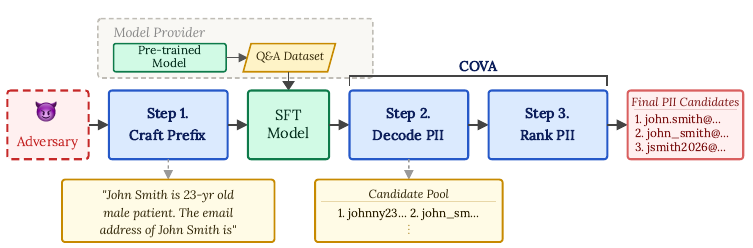}
\caption{Targeted PII reconstruction pipeline. The attacker constructs a prefix from the available target information, searches for plausible PII candidates, and re-ranks the candidate pool using an estimate of the model's generic PII prior.}
\label{fig:attack_pipeline}
\end{figure}

Figure~\ref{fig:attack_pipeline} presents the targeted PII reconstruction pipeline. Given the information available under an attacker-knowledge setting, the attacker first constructs a target prefix $\mathrm{P_t}$ for querying the fine-tuned model. The prefix is crafted directly from the attacker's knowledge of the target and may contain only the target identity or additional information, such as demographics or a redacted version of the original SFT record.

The attacker then applies COVA to explore the model's continuation distribution and construct a pool of plausible PII candidates. As part of decoding procedure, COVA estimates the model's generic preference for each candidate using a target-independent prefix $P_g$. Unlike $P_t$, $P_g$ is not
constructed from the attacker's knowledge of the target; it specifies only the PII type and excludes the target identity and any target-specific information. For example, when reconstructing a date of birth, $P_g$ may be \textit{``The date of birth is''}. COVA then adjusts the candidate ranking
using the likelihood differential between $P_t$ and $P_g$ and returns the final ranked list of reconstructed PII candidates. 

\subsection{Attacker Knowledge and Prefixes}
\label{attacker_knowledge_prefix}

We specify the information available to the attacker and how that information is converted into prefixes for targeted reconstruction. We first describe our prompt-side design and the corresponding baseline, then define the attacker-knowledge settings and the prefix used in each setting.

\noindent\textbf{Prompt-Side Baselines.}
Prior work~\cite{janus-finetuning, pii-extraction-optimized-prompt, few-shot-prompting} on prefix-based PII extraction primarily focuses on prompt-side improvements. For example, Keum et al.~\cite{pii-extraction-optimized-prompt} generate and select prompts intended to elicit large numbers of memorized PII strings, while Cheng et al.~\cite{few-shot-prompting} prepend few-shot PII demonstrations and augment targeted queries with auxiliary information. Cheng et al.~\cite{few-shot-prompting} provide the closest baseline for our targeted PII reconstruction, but their augmentation relies on public PII pairs and target-related auxiliary information that can help recover public PII learned during pre-training. Such information does not reveal associations contained only in a proprietary SFT dataset. We therefore construct each prefix directly from the information available under the corresponding attacker-knowledge setting, without supplementing it with external auxiliary information.
 
\noindent\textbf{Attacker Knowledge Settings.}
We evaluate targeted PII reconstruction under several attacker-knowledge settings. Table~\ref{tab:attacker_knowledge} summarizes the information available in each setting. While the attacker knows the user's name across all settings, each differs in the additional information available for constructing the attack prefix. The redacted-record setting provides the most target-specific information, whereas the name-only setting provides the least. The intermediate settings capture different forms of partial knowledge and are not necessarily strictly ordered by strength.

For settings involving contextual information, we derive two compressed representations of the original non-PII context. A \emph{sentence-level summary} condenses the relevant content into at most two sentences, while a \emph{keyword-level summary} represents the context using fewer than ten words. Because the original records are generally written as full paragraphs, even the sentence-level summaries provide substantially less information than the redacted original record.

\noindent\textbf{Prefix Construction.}
\label{prefix_construction}
For each attacker-knowledge setting, we convert the available information into a prefix intended to elicit the target PII as a continuation. Most prefixes use a common field-completion structure, e.g., \textit{``[Description about Person A]. The [PII field] of Person A is''}, so that differences in reconstruction performance primarily reflect attacker knowledge rather than unrelated prompt-design choices. Detailed construction rules and example prefixes are provided in Appendix~\ref{a_prefix}.

\begin{table}[t!]
\centering
\caption{Attacker-knowledge settings used to construct reconstruction
prefixes. All settings include the target identity.}
\label{tab:attacker_knowledge}
\footnotesize
\small
\begin{tabular}{
p{0.22\columnwidth}
p{0.66\columnwidth}}
\hline
\textbf{Setting} & \textbf{Available Information} \\
\hline

Redacted
& The target SFT record with PII removed. \\

Summary
& A sentence-level summary of the target's contextual information, such as
medical symptoms or legal circumstances. \\

Keywords
& A short keyword-level representation of the target's contextual
information. \\

Demographics
& Demographic and background attributes, such as age, gender, race, or profession. \\

In-distribution
& Knowledge of the dataset domain and response format, represented using
generic domain-specific content without target-specific context. \\

Name
& No additional information beyond the target identity. \\

\hline
\end{tabular}
\end{table}

\subsection{Adversarial Decoding Design}
\label{adversarial_decoding}
Given a target prefix $P_t$, the objective of the decoding stage is to construct a set of plausible PII continuations and return the top-N candidates. 
 
\noindent\textbf{Decoding Baselines.}
While prompt-side methods modify the prefix used to elicit PII, the decoding strategy determines which candidate continuations are explored and retained.
Prior PII-extraction studies primarily use top-$k$ sampling and beam search, which are standard decoding strategies for LLMs~\cite{carlini-extraction,janus-finetuning, few-shot-prompting, pii-extraction-optimized-prompt, analyzing-pii-leakage}. These methods face two limitations: insufficient candidate coverage and bias from the model's generic PII prior.

\noindent\emph{Coverage.}
Top-$k$ sampling restricts each step to a high-probability token set, while beam search retains only the top-$B$ partial sequences by cumulative
log-likelihood. Both methods therefore favor continuations that are locally likely. However, the correct PII may begin with one or more low-probability tokens and consequently may not be sampled, even if the resulting complete sequence would rank highly among the final candidates. 
This is particularly problematic for exact PII reconstruction, where one incorrect
token causes failure and partial-sequence rankings may change as later tokens are generated~\cite{differentiation}. For example, ``12\slash1'' may initially
be more likely than ``12\slash3,'' although the completed sequence "12\slash31" may ultimately outrank "12\slash11."

\noindent\emph{Generic PII Prior.}
Candidate likelihood under an SFT model reflects both target-specific information learned during fine-tuning and generic preferences inherited from
pre-training. This distinction is especially important when fine-tuning uses a small, task-specific dataset: common dates or locations, may remain highly ranked even when they are unrelated to the target. Prior work addresses this issue by comparing likelihoods under the fine-tuned and base models~\cite{differentiation}. However, this requires access to the exact base model, which is unrealistic in practice.

These two limitations motivate COVA's coverage-aware search and prior-differential re-ranking.

\begin{figure}[h]
    \centering
    \includegraphics[width=\linewidth]{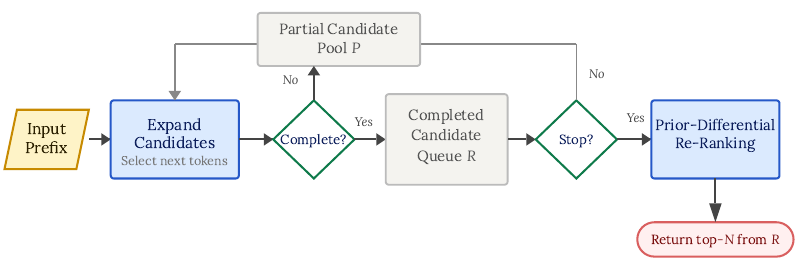}
    \caption{\textbf{Illustration of COVA.} (1) The \emph{Partial Candidate Pool} $\mathcal{P}$ stores active, incomplete candidates. (2) \emph{Expand Candidates} extends each candidate using the adaptive top-$p$/top-$k$ token set. (3) The \emph{Completion Check} determines whether an extension is valid and complete for the target PII format. (4) The \emph{Completed Candidate Queue} $\mathcal{R}$ retains the highest-scoring complete candidates. (5) The \emph{Stop Condition} terminates the search when no partial candidate can surpass the lowest-scoring retained candidate. (6) \emph{Prior-Differential Re-Ranking} adjusts selected candidates using the difference between their target-specific and generic-prefix log-likelihoods.}
    \label{fig:algorithm}
\end{figure}

\noindent \textbf{Our Approach.} 
We propose \textbf{COV}erage-\textbf{A}ware Decoding
(\textbf{COVA}), a novel deterministic adversarial decoding method that combines
coverage-aware candidate search with prior-differential re-ranking. The search stage constructs a diverse set of plausible complete PII strings to improve coverage, and the re-ranking stage reduces the influence of candidates that are inherently likely under the model's generic PII distribution. Figure~\ref{fig:algorithm} illustrates the two integrated components of COVA, and Algorithm~\ref{alg:cova_algorithm} provides the complete procedure.

During candidate search, COVA maintains a global pool of partial sequences (\textbf{Partial Candidate Pool P} in Figure~\ref{fig:algorithm}) and expands every active candidate before performing pruning (\textbf{Expand Candidates} in Figure~\ref{fig:algorithm}). For each partial candidate, COVA deterministically explores next-token options while adapting to the shape of the model's distribution. To control computational cost, COVA constructs both top-$p$ and top-$k$ expansion sets and uses the one that produces fewer candidate branches at each step. For example, DOB generation may benefit more from top-$p$ because probability mass is often concentrated on a few digits. For names or cities, top-$k$ may be more appropriate when the next-token distribution is flatter, and top-$p$ would retain too many plausible tokens. Each extension is scored by cumulative log-likelihood, and invalid continuations are discarded.

Unlike beam search, COVA does not enforce a fixed beam-width constraint after every expansion. Instead, it allows the candidate pool to grow before applying global sequence-level pruning. This gives candidates with lower-probability early tokens additional opportunities to develop into plausible complete PII strings.

A candidate enters the completed-candidate queue once it satisfies the format constraints of the target PII (\textbf{Completion Check} in Figure~\ref{fig:algorithm}). We use regular expressions for structured fields, such as dates of birth and email addresses, and a non-alphabetic boundary for less structured fields. To avoid favoring candidates that complete early, search continues until no remaining partial sequence can exceed the lowest-scoring candidate in $\mathcal{R}$, which is sorted by log-likelihood.

However, the likelihood used during search reflects both information associated with the target and the model's generic prior over the relevant PII type. For example, common dates and locations may receive high likelihood even when they are unrelated to the target individual. COVA therefore supports \textbf{prior-differential re-ranking} (Figure~\ref{fig:algorithm}) of selected completed candidates using a second, target-independent prefix. 

Let $P_t$ denote the target-specific prefix and let $P_g$ denote a generic prefix that specifies the PII type but excludes the target identity and target-specific context. For example, a generic date-of-birth prefix may be \textit{``The date of birth is''}. Let $\ell_f(C \mid P)$ denote the cumulative log-likelihood of candidate $C$ under prefix $P$ using the attacked fine-tuned model. COVA computes

\begin{equation}
s_{\mathrm{COVA}}(C)
=
\ell_f(C \mid P_t)
-
\ell_f(C \mid P_g).
\label{eq:cova-differential}
\end{equation}

This score promotes candidates whose likelihood is specifically increased by the target information and suppresses candidates that are likely regardless of the target. 

\begin{algorithm}[!t]
\caption{COVA}
\label{alg:cova_algorithm}
\footnotesize
\renewcommand{\baselinestretch}{0.94}\selectfont
\begin{algorithmic}[1]
\Require Fine-tuned model $M_f$, tokenizer $\mathcal{T}$, prefix $P$, max tokens $T_{\max}$
\Require $\mathrm{Valid}(\cdot)$, $\mathrm{Complete}(\cdot)$, selector $\mathrm{Select}(\pi,\theta)$
\Require $K_{\text{keep}}$, $K_{\text{prune}}$, $N_{\text{best}}$, $N$
\Require Optional generic prefix $P_g$ and re-ranking size $K_{\text{rerank}}$
\Ensure Top $N$ completed candidates

\State $\mathcal{P}\gets\{(0,\epsilon,0)\}$; $\mathcal{R}\gets\emptyset$

\While{$\mathcal{P}\neq\emptyset$}
    \State $\mathcal{P}_{\mathrm{new}}\gets\emptyset$
    \ForAll{$(s,c,t)\in\mathcal{P}$}
        \If{$t=T_{\max}$}
            \State \textbf{continue}
        \EndIf

        \State $x\gets\mathcal{T}(P\|c)$; $\pi\gets\mathrm{Softmax}(M_f(x))$

        \ForAll{$u\in\mathrm{Select}(\pi,\theta)$}
            \State $c'\gets c\|\mathcal{T}^{-1}(u)$
            \If{$\neg\mathrm{Valid}(c')$}
                \State \textbf{continue}
            \EndIf

            \State $s'\gets s+\log\pi(u)$
            \If{$\mathrm{Complete}(c')$}
                \State Insert $(s',c')$ into $\mathcal{R}$ if among the top $N_{\text{best}}$
            \Else
                \State Add $(s',c',t+1)$ to $\mathcal{P}_{\mathrm{new}}$
            \EndIf
        \EndFor
    \EndFor

    \State $\mathcal{P}\gets\mathcal{P}_{\mathrm{new}}$
    \If{$|\mathcal{P}|\ge K_{\text{keep}}$}
        \State Keep the top $K_{\text{prune}}$ candidates in $\mathcal{P}$
    \EndIf
    \If{$|\mathcal{R}|\ge K_{\text{prune}}$ and
        $\max_{(s,c,t)\in\mathcal{P}}s\le\mathrm{MinScore}(\mathcal{R})$}
        \State \textbf{break}
    \EndIf
\EndWhile

\State Sort $\mathcal{R}$ by score in descending order

\If{$P_g$ is provided}
    \State $\mathcal{R}_{\text{rerank}}\gets$ first $K_{\text{rerank}}$ candidates in $\mathcal{R}$
    \ForAll{$(s,c)\in\mathcal{R}_{\text{rerank}}$}
        \State $s_g\gets\mathrm{LogLik}(M_f,P_g,c)$; $s_{\text{diff}}\gets s-s_g$
        \State Replace the score of $c$ with $s_{\text{diff}}$
    \EndFor
    \State Reorder $\mathcal{R}_{\text{rerank}}$ by $s_{\text{diff}}$ in descending order
    \State Replace the corresponding subset in $\mathcal{R}$ with $\mathcal{R}_{\text{rerank}}$
\EndIf

\State \Return top $N$ candidates in $\mathcal{R}$
\end{algorithmic}
\end{algorithm}

\section{Dataset Creation}
\label{dataset_creation}
In this section, we describe how we construct custom SFT datasets from existing Q\&A data to incorporate PII-dependent tasks. The first dataset is derived from MedRedQA~\cite{medredqa}, a medical Q\&A dataset, and the second from a legal advice dataset~\cite{legal-advice}. Both source datasets originate from Reddit discussions. We will release the resulting datasets to support reproducibility and facilitate future research on PII leakage in SFT models.

\subsection{Motivation}
\label{dataset-motivation}
Our goal is to leverage publicly available datasets to construct realistic Q\&A datasets in sensitive domains that incorporate PII. Existing datasets have several limitations. First, datasets commonly used for PII extraction~\cite{enron, echr, ai4privacy} are unstructured text corpora rather than Q\&A formats. Second, many Q\&A datasets are not user-centric and do not encode attributes specific to an individual user. For example, alignment datasets~\cite{alpaca, self-instruct} and domain-specific datasets for mathematics or coding~\cite{ossinstruct, mathinstruct} are designed to improve instruction-following or technical capabilities but generally lack user-specific context. However, a small number of Q\&A datasets~\cite{legal-advice,medredqa} in sensitive domains (e.g., medical or legal) are derived from real user interactions on public forums, and thus better reflect authentic, user-centered scenarios. Using these datasets as a foundation aligns with our goal of constructing realistic Q\&A data, which would be difficult to replicate with synthetic data.

Beyond these structural limitations, widely used existing PII corpora~\cite{ai4privacy, enron, echr} may already have appeared during pre-training, making it difficult to attribute PII leakage specifically to fine-tuning. Consequently, prior work on fine-tuned models~\cite{pii-extraction-optimized-prompt, hidden-dangers} may primarily capture amplified extraction of previously learned PII rather than leakage introduced by fine-tuning itself.

To address these limitations while preserving semantic fidelity, we retain the original Q\&A samples and extend each single-turn example into a multi-turn conversation by adding PII-related tasks with synthetic PII (e.g., drafting or editing emails). This allows us to leverage high-quality human-generated data while introducing controlled PII through a structured persona framework that enables consistent user identities across samples. We ensure that user PII appears in the assistant's responses. This is because the standard SFT objective applies the training loss to the assistant's response only, rather than the user's instruction. As a result, PII appearing in assistant outputs is directly optimized during training, whereas PII appearing only in user instructions influences the model indirectly through conditioning. 

Prior work has shown that PII duplication has a significant impact on reconstruction success~\cite{analyzing-pii-leakage}. Our work incorporates this insight and investigates how varying levels of duplication affect reconstruction. However, naively duplicating PII within the dataset is insufficient, since doing so risks breaking consistency across user personas. Samples sharing the same PII are treated as belonging to the same user and must agree on attributes such as demographics and medical history. Since the original Q\&A datasets lack explicit user profiles, we introduce an LLM-based annotation stage to recover those attributes and support consistent persona construction. 

\subsection{Dataset Generation Process}
The dataset construction process involves four steps: (1) defining tasks, (2) data annotation, (3) constructing user profiles with synthetic PII, and (4) multi-turn dataset creation. Our goal is to extend high-quality single-turn Q\&A datasets into multi-turn SFT datasets that incorporate PII by adding second- and third-turn exchanges. We use the Llama-3.3-70B Instruct model for dataset generation. To assess generation quality, we manually inspect a sampled subset of the resulting examples throughout the generation process for attribute consistency, PII correctness, and conversational coherence.

\noindent \textbf{Defining Tasks.}
\label{defining_task} 
Starting from an initial user question, we define follow-up tasks in multi-turn conversations that require PII. These include email editing or drafting, questions involving sensitive personal information (e.g., insurance coverage), and location-based queries using retrieval-augmented generation (RAG), such as finding nearby hospitals or clinics. Table~\ref{tab:subsequent_turn_examples} shows several example tasks from each dataset. The full set of tasks is provided in Appendix~\ref{a_dataset_annotation} Table~\ref{tab:subsequent_turn}.

\begin{table}[htbp]
\centering
\footnotesize
\setlength{\tabcolsep}{3pt}
\renewcommand{\arraystretch}{1.05}
\caption{Example task definitions for multi-turn interactions in the medical
and legal datasets.}
\label{tab:subsequent_turn_examples}

\begin{tabularx}{\columnwidth}{
>{\raggedright\arraybackslash}p{0.20\columnwidth}
>{\raggedright\arraybackslash}X
>{\raggedright\arraybackslash}p{0.26\columnwidth}}
\toprule
\textbf{Task Type} & \textbf{Description} & \textbf{User PII} \\
\midrule

\multicolumn{3}{c}{\textbf{Medical Q\&A}} \\
\midrule

HOSPITAL
& Find nearby hospitals or clinics using the user's location via RAG.
& Location \\

EDIT EMAIL
& Edit the user's draft email to their doctor.
& Name, DOB, email, phone \\

INSURANCE QUESTION
& Answer a question about the user's insurance coverage.
& Insurer, insurance type, policy no. \\

\midrule
\multicolumn{3}{c}{\textbf{Legal Q\&A}} \\
\midrule

LAW FIRM
& Find nearby lawyers or law firms using the user's location via RAG.
& Location \\

COURT FILING
& Draft a petition or court filing regarding the user's dispute.
& Name, DOB, address, SSN \\

DOCUMENT REVIEW
& Answer a question about a provided legal-document excerpt.
& Case no. \\

\bottomrule
\end{tabularx}
\end{table}

\noindent \textbf{Data Annotation.}
We extract relevant information from each sample in the original Q\&A dataset, including user attributes (e.g., demographics, physical traits, habits) and the nature of the interaction. This serves two purposes. First, it determines the appropriate follow-up task. For example, in a medical Q\&A dataset, if the original query concerns medication usage, a natural follow-up task would be to locate nearby pharmacies. In contrast, if the user mentions past appointments or procedures, an appropriate follow-up task might involve insurance-related questions or writing follow-up emails to their doctors. For medical datasets, relevant information includes procedure or test status and symptom urgency. For legal datasets, it includes filing status, opposing-party details, and dispute type.

The second objective is to identify and extract user-specific attributes. As described in~\ref{dataset-motivation}, we study how PII duplication affects reconstruction success, and our design ensures PII duplication while maintaining consistency across user profiles. While the original dataset does not contain explicit PII, it often includes references to user characteristics such as demographics (age, gender, race) or background information (e.g., medical history, height, weight, or substance use). These attributes are used to construct consistent user personas so that later, during the persona construction stage, we ensure that any user-related attributes in each sample that belong to the same persona are consistent. For example, a sample describing a female European patient who smokes should not be assigned to a persona characterized as a non-smoking Asian male. An example of how user attributes are extracted from the original sample is included in Appendix~\ref{a_dataset_annotation} Figure~\ref{fig:data-annotation-example}. 

\noindent \textbf{Construction of User Profiles with PII.}
Based on the annotations, we construct a set of user personas with well-defined characteristics. Each sample from the original dataset is assigned to a persona such that the user attributes present in the sample are consistent with the assigned persona. For certain fine-grained attributes, such as height and weight, we relax exact consistency during persona assignment. These attributes are matched using coarse buckets (e.g., 170--175 cm), after which we select a representative persona value and edit conflicting references during post-processing. Appendix~\ref{a_pii} (Table~\ref{tab:user_data_clean} and~\ref{tab:pii_generation}) provides details on the attributes and PII associated with each persona and methods used to generate each synthetic PII field. We assume that users reside in U.S. cities~\cite{uscities}, which is broadly consistent with the fact that the majority of Reddit users are based in the United States, as both Q\&A datasets are originally scraped from Reddit forums (with a small portion of legal samples assigned to Canada).

\begin{figure}[htbp]
    \centering
    \includegraphics[width=0.95\linewidth]{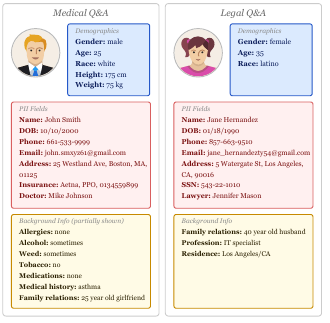}
    \caption{Example of User Profile with Synthetic PII}
    \label{fig:user_profile_example}
\end{figure}

\noindent \textbf{Multi-turn Dataset Creation.}
For each sample, we generate additional conversation turns based on the assigned task and the persona's PII. These subsequent turns simulate realistic interactions between the user and assistant for tasks requiring the use of user PII. We use few-shot prompting with an LLM to generate second and third turns for each sample based on the assigned task and the persona's PII. In particular, we provide the model with the task objective, the initial conversation turn, and the relevant PII, and prompt it to craft the user query and the corresponding assistant's response. For tasks requiring retrieval-augmented generation (RAG), we prepend relevant retrieved content to the user query using publicly available datasets~\cite{hospitals, hrsa_data_warehouse, gigasheet_law_firms}. For example, in location-based queries such as identifying nearby medical facilities, we filter these datasets by the user's ZIP code to construct a JSON-formatted list of real-world entities; this list is prepended as context directly before the user's question. An example training sample is illustrated in Appendix~\ref{a_training_sample}, while PII duplication statistics are provided in Appendix~\ref{a_dataset_statistics}.
\section{Evaluation}
\label{evaluation}
In this section, we evaluate targeted PII reconstruction from models trained on our medical and legal SFT datasets. We first compare COVA with prompt-side and decoding baselines, and then examine how reconstruction varies with attacker knowledge, model scale, and model family. Finally, we evaluate data scrubbing as a defense and measure its impact on downstream model utility.

\subsection{Experiment Setting}
\noindent \textbf{Model Setup.}
We use the Qwen-2.5 7B model obtained from the Hugging Face Hub as our primary model, and we further evaluate our approach across models of varying sizes and families. For finetuning, we use 90\% of the dataset for training and 10\% for validation, and fine-tune for 10 epochs, selecting the checkpoint at epoch 5 based on the lowest validation loss. Details of the training setup are provided in Appendix~\ref{a_training_setup}. In addition, we conduct the same extraction attacks on models fine-tuned from instruction-tuned (aligned) models to evaluate the impact of prior alignment on leakage behavior, and these results are reported in Appendix~\ref{a_instruct}.

\noindent\textbf{Attack Configuration.}
We evaluate reconstruction of three PII attributes present in both the medical and legal datasets: date of birth (DOB), email address, and city of residence. Email and location are commonly examined in prior PII leakage studies~\cite{analyzing-pii-leakage, janus-finetuning, pii-extraction-optimized-prompt, few-shot-prompting}, while DOB extends the evaluation to a higher-entropy PII type with substantially lower inherent predictability. For COVA~\ref{alg:cova_algorithm}, we apply prior-differential re-ranking to the top 20 candidates for DOB reconstruction and the top 10 candidates for city reconstruction. We omit prior-differential re-ranking for email because a generic prefix such as \textit{"The email is"} does not provide a meaningful prior when email addresses depend strongly on the target name. The remaining COVA parameters are provided in Appendix~\ref{a_cova}. All experiments are conducted on three H100 GPUs.

\noindent \textbf{Reconstruction Metrics.}
We evaluate reconstruction using Recall@$N$, defined as the fraction of attempts for which the correct PII appears among the top-$N$ returned candidates. We report Recall@5, Recall@25, and Recall@50 to capture both near-top reconstruction and broader candidate coverage.

Different PII types have different levels of inherent guessability, and different base models may assign substantially different prior probabilities to the same PII. To account for these effects when analyzing attacker-knowledge settings or comparing across models, we evaluate both the base and fine-tuned checkpoints and report $\Delta=\mathrm{FT}-\mathrm{Base}$. This differential captures the increase in reconstruction attributable to SFT beyond what can already be recovered from the base model.

For city reconstruction, we exclude New York City and Los Angeles because the base model predicts these cities for nearly all prompts even before fine-tuning. Their inclusion would therefore inflate reconstruction rates through trivial prior-based matches rather than leakage introduced by the SFT data.

\subsection{Comparison with Baseline Methods}
We compare COVA with prompt-side and decoding baselines. Prompt-side methods modify or augment the prefix used to elicit the target PII, whereas decoding-side methods hold the prefix fixed and change how candidate continuations are sampled, explored, and ranked.

\noindent\textbf{Prompt-Side Baseline Comparison.}
\label{eval:prompt_side_baseline}
We hypothesize that few-shot demonstrations constructed from public PII pairs (Cheng et al.~\cite{few-shot-prompting}) will not transfer effectively to private associations introduced through proprietary SFT data. To test this hypothesis in the name-only attacker-knowledge setting, we prepend eight public identity--PII pairs as demonstrations before the target query while keeping the target information and decoding method unchanged. For email reconstruction, we obtain public name--email pairs from the Enron dataset~\cite{enron}. For city and DOB reconstruction, we construct name--attribute pairs for public figures using Wikidata~\cite{wikidata}. None of the demonstrations are drawn from our private SFT datasets.

\begin{table}[t]
\centering
\footnotesize
\setlength{\tabcolsep}{4pt}
\renewcommand{\arraystretch}{0.95}
\caption{Reconstruction performance (\%) on Qwen2.5-7B (medical dataset, name-only knowledge). ``P'' denotes the original prefix, while ``FS'' prepends eight target--PII pair demonstrations. Few-shot examples do not reliably improve reconstruction across PII types and decoding methods.}
\label{tab:fewshot_comparison}

\begin{tabular}{@{}ll*{3}{cc}@{}}
\toprule
& &
\multicolumn{2}{c}{Top-$k$} &
\multicolumn{2}{c}{Beam} &
\multicolumn{2}{c}{COVA} \\
\cmidrule(lr){3-4}
\cmidrule(lr){5-6}
\cmidrule(lr){7-8}

PII & Threshold.
& P & FS
& P & FS
& P & FS \\
\midrule

\multirow{3}{*}{City}
& Recall@5
& \textbf{1.31} & 0.15
& \textbf{0.29} & 0.15
& \textbf{2.48} & 1.46 \\

& Recall@25
& \textbf{5.39} & 1.60
& \textbf{5.54} & 1.46
& \textbf{9.18} & 4.96 \\

& Recall@50
& \textbf{8.60} & 2.33
& \textbf{6.85} & 1.46
& \textbf{15.74} & 4.96 \\

\cmidrule(lr){1-8}

\multirow{3}{*}{DOB}
& Recall@5
& 0.00 & 0.00
& 0.00 & 0.00
& \textbf{0.10} & 0.00 \\

& Recall@25
& 0.00 & 0.00
& 0.00 & \textbf{0.10}
& 0.10 & \textbf{0.29} \\

& Recall@50
& 0.00 & 0.00
& 0.00 & \textbf{0.19}
& 0.19 & \textbf{0.58} \\

\cmidrule(lr){1-8}

\multirow{3}{*}{Email}
& Recall@5
& \textbf{1.93} & 1.75
& \textbf{4.03} & 1.31
& \textbf{4.64} & 1.40 \\

& Recall@25
& 3.59 & \textbf{8.93}
& \textbf{9.98} & 5.08
& \textbf{14.10} & 6.48 \\

& Recall@50
& 4.03 & \textbf{10.42}
& \textbf{10.07} & 5.43
& \textbf{15.85} & 7.97 \\

\bottomrule
\end{tabular}
\end{table}
Table~\ref{tab:fewshot_comparison} compares reconstruction with and without public few-shot demonstrations across three decoding methods for the medical dataset. The results on the legal dataset are reported in Appendix~\ref{appendix_prompt_side}. Few-shot prompting does not provide a consistent improvement. For city reconstruction, it substantially reduces performance across all methods: for example, COVA Recall@50 decreases from 15.74\% to 4.96\%. Few-shot prompting also reduces email reconstruction for beam search and COVA: Recall@50 decreases from 10.07\% to 5.43\% for beam search and from 15.85\% to 7.97\% for COVA. The main exception is top-$k$ sampling, for which Recall@50 increases from 4.03\% to 10.42\%. DOB reconstruction remains close to zero overall, with only small absolute improvements at higher thresholds. Overall, public few-shot demonstrations do not reliably improve recovery of private SFT-only associations; in many cases, they reduce performance. 

\begin{findingbox}
\textbf{Finding:} Public few-shot demonstrations do not reliably improve reconstruction of private SFT-only PII associations and reduce performance for many PII types and decoding methods.
\end{findingbox}

\begin{table*}[t!]
\centering
\small
\setlength{\tabcolsep}{1.6pt}
\caption{Reconstruction performance (\%) on Qwen2.5-7B across decoding strategies and datasets. Redacted represents the strongest attacker-knowledge setting, whereas Name represents the weakest. COVA outperforms the decoding baselines across most PII types and attacker-knowledge settings. The results on Llama~3.1~8B are reported in Appendix Table~\ref{tab:decoding_compare_llama}.}

\label{tab:decoding_compare}
\footnotesize

\begin{tabular}{@{}ll*{6}{ccc}@{}}
\toprule
& &
\multicolumn{9}{c}{\textbf{Medical}}
& \multicolumn{9}{c}{\textbf{Legal}} \\
\cmidrule(lr){3-11}
\cmidrule(lr){12-20}

& &
\multicolumn{3}{c}{City}
& \multicolumn{3}{c}{Email}
& \multicolumn{3}{c}{DOB}
& \multicolumn{3}{c}{City}
& \multicolumn{3}{c}{Email}
& \multicolumn{3}{c}{DOB} \\
\cmidrule(lr){3-5}
\cmidrule(lr){6-8}
\cmidrule(lr){9-11}
\cmidrule(lr){12-14}
\cmidrule(lr){15-17}
\cmidrule(lr){18-20}

Knowledge & Thres.
& Top-$k$ & Beam & COVA
& Top-$k$ & Beam & COVA
& Top-$k$ & Beam & COVA
& Top-$k$ & Beam & COVA
& Top-$k$ & Beam & COVA
& Top-$k$ & Beam & COVA \\
\midrule

\multirow{3}{*}{Redacted}
& Recall@5
& 1.35 & 2.63 & \textbf{3.74}
& 3.22 & \textbf{7.90} & 7.30
& 0.10 & 0.10 & \textbf{0.25}
& 1.93 & 4.87 & \textbf{8.27}
& 5.80 & 7.66 & \textbf{7.87}
& \textbf{0.21} & 0.00 & 0.10 \\

& Recall@25
& 4.87 & 10.43 & \textbf{12.65}
& 5.99 & 17.77 & \textbf{19.22}
& 0.25 & 0.71 & \textbf{0.96}
& 8.27 & 16.42 & \textbf{26.61 }
& 8.90 & 17.39 & \textbf{18.43}
& 0.31 & \textbf{1.04} & 0.93\\

& Recall@50
& 9.10 & 16.50 & \textbf{23.48}
& 6.24 & 17.77 & \textbf{19.93}
& 0.25 & 1.31 & \textbf{2.47}
& 14.84 & 21.29 & \textbf{35.45}
& 8.90 & 17.70 & \textbf{20.08}
& 0.31 & \textbf{1.45} & \textbf{1.45} \\
\cmidrule(lr){1-20}

\multirow{3}{*}{Summary}
& Recall@5
& 0.51 & 0.20 & \textbf{2.94}
& 2.94 & 6.58 & \textbf{7.59}
& 0.00 & 0.00 & \textbf{0.10}
& 0.79 & 1.93 & \textbf{6.80}
& 2.38 & 4.55 & \textbf{5.69}
& \textbf{0.10} & 0.00 & \textbf{0.10} \\

& Recall@25
& 4.96 & 6.38 & \textbf{9.11}
& 4.96 & 15.38 & \textbf{18.62}
& 0.00 & \textbf{0.30} & 0.20
& 6.91 & 21.06 & \textbf{25.71}
& 3.52 & 10.77 & \textbf{14.39}
&0.10 & 0.31 & \textbf{0.52} \\

& Recall@50
& 8.30 & 9.41 & \textbf{19.93}
& 4.98 & 15.38 & \textbf{20.95}
& 0.00 & 0.30 & \textbf{0.61}
& 13.25 & 26.27 & \textbf{36.35}
& 3.52 & 10.87 & \textbf{17.49}
& 0.10 & \textbf{0.72} & \textbf{0.72} \\
\cmidrule(lr){1-20}

\multirow{3}{*}{Name}
& Recall@5
& 1.31 & 0.29 & \textbf{2.48}
& 1.93 & 4.03 & \textbf{4.64}
& 0.00 & 0.00 & \textbf{0.10}
& 0.44 & 0.00 & \textbf{5.09}
& 2.98 & 4.79 & \textbf{5.12}
& 0.00 & 0.00 & 0.00 \\

& Recall@25
& 5.39 & 5.54 & \textbf{9.18}
& 3.59 & 9.98 & \textbf{14.10}
& 0.00 & 0.00 & \textbf{0.10}
& 7.57 & 6.70 & \textbf{30.42}
& 4.71 & 10.58 & \textbf{12.73}
& 0.00 & 0.00 & \textbf{0.24} \\

& Recall@50
& 8.60 & 6.85 & \textbf{15.74}
& 4.03 & 10.07 & \textbf{15.85}
& 0.00 & 0.00 & \textbf{0.19}
& 15.87 & 6.70 &\textbf{ 35.95}
& 5.12 & 10.58 & \textbf{14.88}
& 0.00 & 0.00 & \textbf{0.39} \\

\bottomrule
\end{tabular}
\end{table*}

\noindent \textbf{Decoding-side Baseline Comparison.}
\label{eval:decoding_baseline}
We evaluate COVA against standard decoding methods used in prior PII leakage studies: top-$k$ sampling~\cite{carlini-extraction, analyzing-pii-leakage, pii-extraction-optimized-prompt} and beam search~\cite{janus-finetuning}. We consider three attacker-knowledge settings defined in Table~\ref{tab:attacker_knowledge}: redacted data, symptom summary, and name only. Table~\ref{tab:decoding_compare} reports reconstruction performance on both datasets using the Qwen2.5-7B model. Appendix~\ref{a_decoding} Table~\ref{tab:decoding_compare_llama} additionally reports Llama-3-8B results, demonstrating COVA's effectiveness across different model families. 

For top-$k$ sampling, we set $k=40$, following prior work~\cite{pii-extraction-optimized-prompt,analyzing-pii-leakage}, and repeatedly sample until the candidate set contains unique values or the retry limit is reached. For beam search, we use a beam width of $B=100$.

COVA generally provides the strongest reconstruction performance across PII types and attacker-knowledge settings, with the largest gains observed for city reconstruction. At Recall@50 for the medical dataset, COVA improves city reconstruction over the strongest baseline from 16.50\% to 23.48\% under redacted data, from 9.41\% to 19.93\% under symptom summaries, and from 8.60\% to 15.74\% under name-only knowledge. Even at the restrictive Recall@5, COVA  improves from 2.63\% to 3.74\%, 0.51\% to 2.94\%, and 1.31\% to 2.48\% across the same settings. City names vary in token length, and the model’s prior is concentrated on a small set of frequently generated locations. This makes city reconstruction particularly sensitive to early token-level pruning and prior-driven rankings.

Across attacker-knowledge settings, COVA's advantage is more pronounced when less target-specific information is available (e.g., summaries, name). In the legal dataset, COVA improves email reconstruction at Recall@50 from 10.87\% to 17.49\% under symptom-summary knowledge, compared with 17.70\% to 20.08\% under redacted data. A similar pattern appears for city reconstruction at Recall@5: COVA improves performance from 0.44\% to 5.09\% under name-only knowledge, compared with 4.87\% to 8.27\% under redacted data. 

DOB reconstruction remains low across all decoding methods, particularly under the symptom-summary and name-only settings. Consequently, the absolute differences between methods are small, although COVA generally provides the highest Recall@50.

\begin{findingbox}
    \textbf{Finding:} COVA generally achieves the strongest reconstruction
    performance across attacker-knowledge settings, with the largest gains observed for city reconstruction. Its advantage is particularly
    clear when the attacker has minimal knowledge about the target.
\end{findingbox}

\subsection{Targeted PII Reconstruction by Attacker Knowledge}
\label{attack_result}
We evaluate how reconstruction success changes across attacker-knowledge settings. The adversary aims to reconstruct a user’s city of residence, email address, and date of birth (DOB) associated with a given identity. We evaluate reconstruction performance under varying levels of attacker knowledge as detailed in Table~\ref{tab:attacker_knowledge} using COVA. Because the number of reconstruction attempts per persona can differ across these settings, we first compute the reconstruction success rate for each persona and then average across personas.

\begin{table*}[htbp]
\centering
\small
\setlength{\tabcolsep}{1.8pt}
\caption{Reconstruction rate (\%) on Qwen2.5-7B across attacker-knowledge settings. $\Delta$ denotes the increase in PII reconstruction after fine-tuning (FT $-$ Base). Fine-tuning increases PII reconstruction over the base model in nearly every setting.}
\label{tab:pii_results}
\footnotesize

\begin{tabular}{lll*{6}{ccc}}
\toprule
& &
& \multicolumn{3}{c}{Redacted}
& \multicolumn{3}{c}{In-distribution}
& \multicolumn{3}{c}{Summary}
& \multicolumn{3}{c}{Keywords}
& \multicolumn{3}{c}{Demographics}
& \multicolumn{3}{c}{Name} \\
\cmidrule(lr){4-6}
\cmidrule(lr){7-9}
\cmidrule(lr){10-12}
\cmidrule(lr){13-15}
\cmidrule(lr){16-18}
\cmidrule(lr){19-21}

Dataset & PII & Thres.
& Base & FT & $\Delta$
& Base & FT & $\Delta$
& Base & FT & $\Delta$
& Base & FT & $\Delta$
& Base & FT & $\Delta$
& Base & FT & $\Delta$ \\
\midrule

\multirow{9}{*}{Medical}

& \multirow{3}{*}{City}
& Recall@5
& 0.92 & 3.33 & \textbf{2.41}
& 1.31 & 2.77 & \textbf{1.46}
& 0.05 & 2.31 & \textbf{2.26}
& 0.34 & 2.62 & \textbf{2.28}
& 0.44 & 2.62 & \textbf{2.19}
& 0.44 & 2.48 & \textbf{2.04} \\

& & Recall@25
& 5.99 & 11.09 & \textbf{5.10}
& 6.85 & 11.22 & \textbf{4.37}
& 6.05 & 8.27 & \textbf{2.22}
& 7.07 & 9.45 & \textbf{2.38}
& 7.14 & 9.91 & \textbf{2.77}
& 7.14 & 9.18 & \textbf{2.04} \\

& & Recall@50
& 8.03 & 22.63 & \textbf{14.60}
& 7.58 & 22.30 & \textbf{14.72}
& 12.06 & 20.00 & \textbf{7.93}
& 12.22 & 19.79 & \textbf{7.57}
& 12.39 & 20.85 & \textbf{8.45}
& 11.08 & 15.74 & \textbf{4.66} \\
\cmidrule(lr){2-21}

& \multirow{3}{*}{Email}
& Recall@5
& 0.74 & 7.43 & \textbf{6.69}
& 2.28 & 7.01 & \textbf{4.73}
& 2.66 & 7.07 & \textbf{4.41}
& 3.14 & 6.60 & \textbf{3.46}
& 2.89 & 7.09 & \textbf{4.20}
& 2.54 & 4.64 & \textbf{2.10} \\

& & Recall@25
& 5.32 & 20.07 & \textbf{14.76}
& 6.39 & 19.88 & \textbf{13.49}
& 7.91 & 17.96 & \textbf{10.05}
& 8.03 & 18.83 & \textbf{10.80}
& 9.37 & 19.70 & \textbf{10.33}
& 6.57 & 14.10 & \textbf{7.53} \\

& & Recall@50
& 8.97 & 20.78 & \textbf{11.81}
& 11.21 & 21.10 & \textbf{9.89}
& 10.75 & 20.73 & \textbf{9.98}
& 11.27 & 21.02 & \textbf{9.75}
& 13.75 & 21.10 & \textbf{7.36}
& 7.88 & 15.85 & \textbf{7.97} \\

\cmidrule(lr){2-21}

& \multirow{3}{*}{DOB}
& Recall@5
& 0.00 & 0.28 & \textbf{0.28}
& 0.00 & 0.00 & \textbf{0.00}
& 0.00 & 0.07 & \textbf{0.07}
& 0.00 & 0.07 & \textbf{0.07}
& 0.00 & 0.19 & \textbf{0.19}
& 0.00 & 0.10 & \textbf{0.10} \\

& & Recall@25
& 0.00 & 0.97 & \textbf{0.97}
& 0.00 & 0.00 & \textbf{0.00}
& 0.15 & 0.15 & \textbf{0.00}
& 0.15 & 0.07 & \textbf{-0.07}
& 0.00 & 0.68 & \textbf{0.68}
& 0.00 & 0.10 & \textbf{0.10} \\

& & Recall@50
& 0.00 & 2.56 & \textbf{2.56}
& 0.00 & 0.19 & \textbf{0.19}
& 0.32 & 0.44 & \textbf{0.12}
& 0.27 & 0.35 & \textbf{0.09}
& 0.00 & 1.46 & \textbf{1.46}
& 0.00 & 0.19 & \textbf{0.19} \\
\midrule

\multirow{9}{*}{Legal}

& \multirow{3}{*}{City}
& Recall@5
& 6.95 & 8.70 & \textbf{1.74}
& 4.51 & 7.13 & \textbf{2.62}
& 4.33 & 5.84 & \textbf{1.50}
& 4.52 & 6.20 & \textbf{1.68}
& 5.24 & 6.99 & \textbf{1.75}
& 3.34 & 5.09 & \textbf{1.74} \\

& & Recall@25
& 25.22 & 26.05 & \textbf{0.83}
& 29.11 & 32.17 & \textbf{3.06}
& 24.87 & 26.09 & \textbf{1.23}
& 24.59 & 25.88 & \textbf{1.30}
& 27.95 & 32.02 & \textbf{4.08}
& 26.45 & 30.38 & \textbf{3.92} \\

& & Recall@50
& 30.66 & 34.12 & \textbf{3.46}
& 37.12 & 44.40 & \textbf{7.28}
& 31.55 & 35.94 & \textbf{4.39}
& 32.57 & 34.17 & \textbf{1.61}
& 35.37 & 44.83 & \textbf{9.46}
& 30.38 & 35.90 & \textbf{5.52} \\
\cmidrule(lr){2-21}

& \multirow{3}{*}{Email}
& Recall@5
& 0.65 & 5.24 & \textbf{4.59}
& 1.16 & 6.94 & \textbf{5.79}
& 0.83 & 3.59 & \textbf{2.77}
& 1.20 & 3.02 & \textbf{1.82}
& 2.07 & 6.20 & \textbf{4.13}
& 1.82 & 5.12 & \textbf{3.31} \\

& & Recall@25
& 3.11 & 12.65 & \textbf{9.54}
& 6.28 & 17.93 & \textbf{11.65}
& 3.59 & 9.71 & \textbf{6.12}
& 4.27 & 10.76 & \textbf{6.49}
& 7.69 & 17.02 & \textbf{9.34}
& 5.04 & 12.73 & \textbf{7.69} \\

& & Recall@50
& 5.73 & 13.55 & \textbf{7.82}
& 10.58 & 20.25 & \textbf{9.67}
& 6.03 & 11.71 & \textbf{5.68}
& 6.57 & 12.38 & \textbf{5.80}
& 11.74 & 19.09 & \textbf{7.36}
& 6.12 & 14.88 & \textbf{8.76} \\

\cmidrule(lr){2-21}

& \multirow{3}{*}{DOB}
& Recall@5
& 0.00 & 0.07 & \textbf{0.07}
& 0.00 & 0.24 & \textbf{0.24}
& 0.00 & 0.15 & \textbf{0.15}
& 0.00 & 0.00 & \textbf{0.00}
& 0.00 & 0.00 & \textbf{0.00}
& 0.00 & 0.00 & \textbf{0.00} \\

& & Recall@25
& 0.15 & 0.91 & \textbf{0.76}
& 0.08 & 0.79 & \textbf{0.71}
& 0.05 & 0.59 & \textbf{0.54}
& 0.00 & 0.29 & \textbf{0.29}
& 0.00 & 0.00 & \textbf{0.00}
& 0.16 & 0.24 & \textbf{0.08} \\

& & Recall@50
& 0.29 & 1.23 & \textbf{0.93}
& 0.16 & 1.65 & \textbf{1.50}
& 0.25 & 0.70 & \textbf{0.45}
& 0.20 & 0.93 & \textbf{0.74}
& 0.00 & 0.00 & \textbf{0.00}
& 0.16 & 0.39 & \textbf{0.24} \\

\bottomrule
\end{tabular}
\end{table*}

Table~\ref{tab:pii_results} reports reconstruction results across the medical and legal Q\&A datasets at Recall@$5$, Recall@$25$, and Recall@$50$. We also compute reconstruction success for the base model before fine-tuning to distinguish SFT-induced leakage from the inherent predictability of each PII type.

Fine-tuning increases reconstruction in nearly all settings, although the magnitude varies across PII types and attacker-knowledge settings. Across both datasets, email exhibits consistently substantial gains. At Recall@5, the increase after fine-tuning ranges from 2.10\% to 6.69\% in the medical dataset and from 1.82\% to 5.79\% in the legal dataset. 

While city reconstruction exhibits similarly substantial gains in the medical dataset, ranging from 4.66\% to 14.72\% at Recall@50, the gains are generally smaller in the legal dataset despite its higher absolute reconstruction rates. The city distribution is more concentrated in the legal dataset, making common cities easier to predict before fine-tuning and producing higher reconstruction for both the base and fine-tuned models. Consequently, the FT--Base difference attributable to SFT is generally smaller in the legal dataset.

\begin{findingbox}
\textbf{Finding:} Fine-tuning increases PII reconstruction over the base model in nearly every setting. The FT--Base gap isolates PII learned during SFT beyond the base model's generic PII predictability,
providing evidence of SFT-induced memorization.
\end{findingbox}

\begin{figure}[h]
    \centering
    \includegraphics[width=\linewidth]{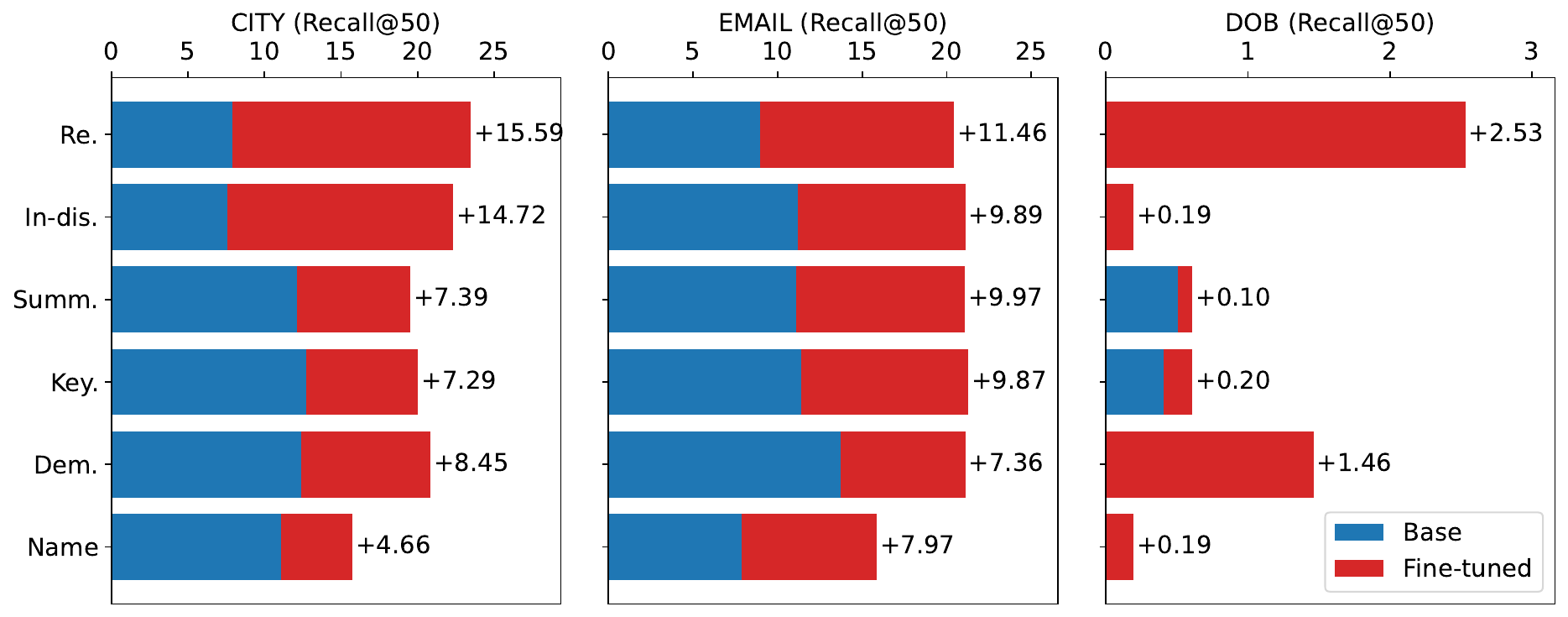}
   \caption{Reconstruction performance (\%) across attacker-knowledge settings for the medical dataset at Recall@50. Labels indicate the change after fine-tuning (FT $-$ Base). Re., In-dis., Summ., Key., and Dem. denote redacted, in-distribution, summary, keywords, and demographics, respectively. Additional information beyond the target name can improve reconstruction gain, while the effect varies by PII type.}
    \label{fig:net_gain_comparison_attack_1}
\end{figure}

To visualize the effect of varying attacker knowledge on reconstruction success,  Figure~\ref{fig:net_gain_comparison_attack_1} visualizes representative medical-dataset results. At Recall@50, the increase in city reconstruction is 15.59\% under the redacted setting and 14.72\% under the in-distribution setting, compared with 4.66\% under name-only knowledge. The summary, keyword, and demographics settings produce intermediate increases of 7.57\%--8.45\%. Email follows the same broad pattern, although the differences between settings are narrower.

DOB reconstruction remains low in absolute terms across both datasets, so the differences among attacker-knowledge settings are correspondingly small. Nevertheless, in the medical dataset, the redacted setting, which provides the strongest attacker knowledge, yields a clearly higher reconstruction rate than most other settings. The main exception is the demographics setting, where providing age-related information improves the model's DOB guesses.

\begin{findingbox}
\textbf{Finding:} Partial attacker knowledge about the target can improve reconstruction of memorized PII associations, even without access to the original training-record prefix.
\end{findingbox}

\subsection{Model Scale and Family}
Finally, we study the effects of model scale and model family on PII leakage. Additional experiments examining the effect of training-data duplication are provided in Appendix~\ref{a_experiment_result}.

\noindent \textbf{Impact of Model Scale.}
Following prior work on PII leakage~\cite{analyzing-pii-leakage,janus-finetuning}, we examine how model scale affects targeted PII reconstruction. We restrict this analysis to the Qwen2.5 family and evaluate 3B, 7B, and 14B models under the symptom-summary attacker-knowledge setting. All models are trained and attacked using the same experimental configuration. 

\begin{figure}[htbp]
    \centering
    \includegraphics[width=1.0\linewidth]{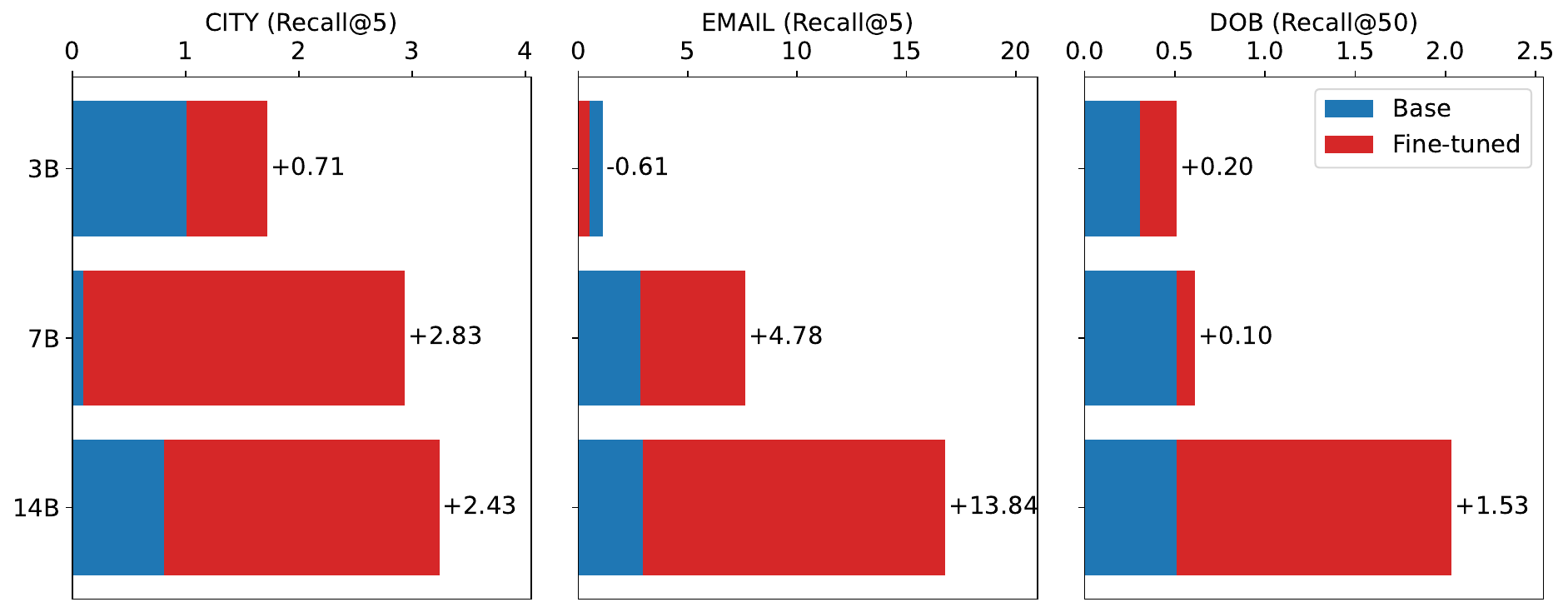}
    \caption{Reconstruction performance (\%) across Qwen2.5 model sizes. City and email are reported at Recall@5, while DOB is reported at Recall@50. Labels show the change after fine-tuning relative to each model's base checkpoint.}
    \label{fig:model_size}
\end{figure}

Figure~\ref{fig:model_size} compares reconstruction before and after fine-tuning, accounting for differences in the prior distributions of the three base models. The additional reconstruction attributable to fine-tuning generally increases with model scale. The clearest trend occurs for email, where Recall@5 changes by \(-0.61\%\), 4.78\%, and 13.84\% for the 3B, 7B, and 14B models, respectively. DOB reconstruction remains low overall, but the 14B model improves by 1.53\%, while the gains for the 3B and 7B models are comparatively small. These results indicate that the largest model exhibits the greatest overall SFT-induced privacy leakage.

\noindent \textbf{Impact of Model Family.}
We next evaluate whether targeted PII reconstruction varies across model families. We compare Qwen2.5-7B with Llama-3.1-8B and DeepSeek-LLM-7B, which are broadly comparable in parameter count. Each model is fine-tuned and evaluated using the same training configuration and symptom-summary attacker-knowledge setting. This comparison isolates differences associated with model family while approximately controlling for scale.  

\begin{figure}[htbp]
    \centering
    \includegraphics[width=1.0\linewidth]{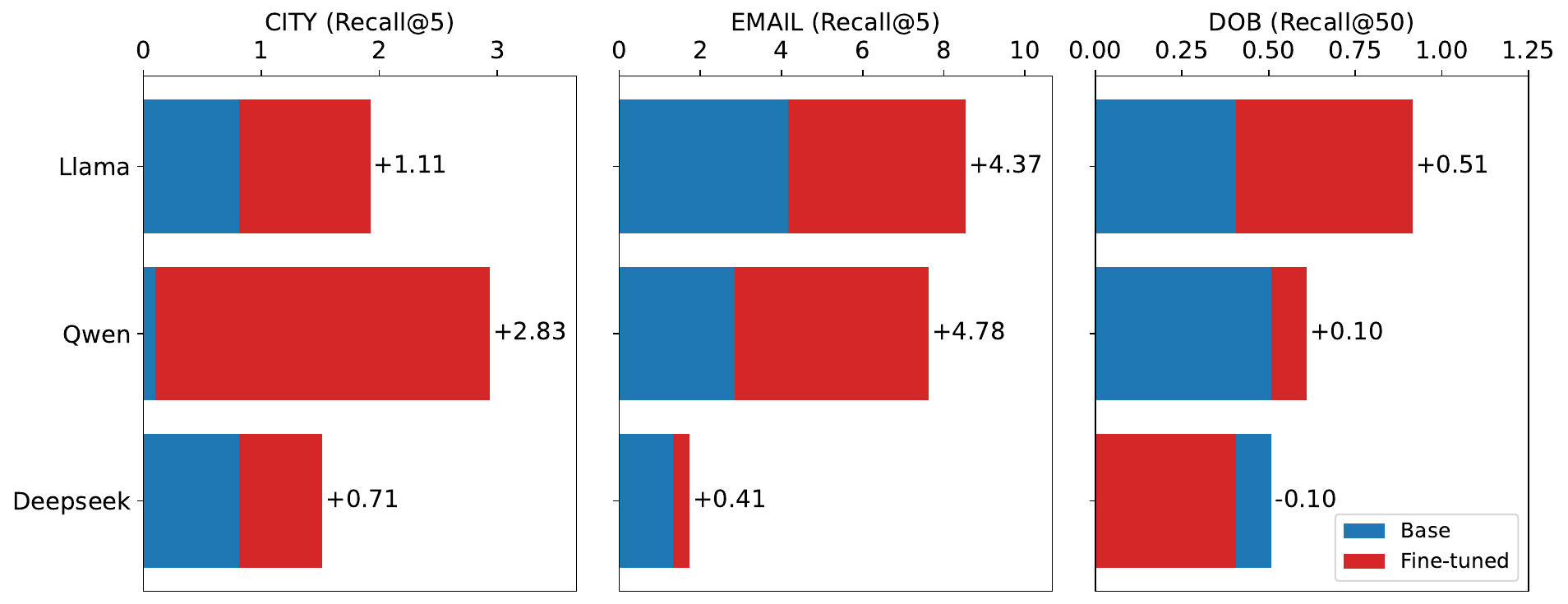}
    \caption{Reconstruction performance (\%) across model families. City and email are reported at Recall@5, while DOB is reported at Recall@50. Labels show the change after fine-tuning relative to each model's base checkpoint.}
    \label{fig:model_family}
\end{figure}

Figure~\ref{fig:model_family} shows that Qwen and Llama generally achieve substantially higher reconstruction success than DeepSeek, which shows only small gains. This difference is particularly clear for email, where Qwen and Llama achieve similar reconstruction rates (4.37--4.78\%) that are much higher than DeepSeek. Between Qwen and Llama, Llama achieves noticeably higher DOB reconstruction, whereas Qwen exhibits higher city leakage. The base-model results also show that model families have different prior distributions over PII types, particularly for city and email, while their DOB priors are more similar. Overall, these results show that PII leakage varies across model families even at comparable parameter counts.

\noindent \textbf{API-Access Model.}
Finally, we investigate targeted PII reconstruction from a closed model accessible only through an API. We fine-tune Gemini-2.5-Flash for five epochs and use COVA to reconstruct city and email under the minimal attacker-knowledge setting, where the adversary knows only the target name. The Gemini API exposes log probabilities for only the top 20 next-token candidates at each decoding step and does not support scoring a specified completed candidate under an alternative prefix. We therefore adapt COVA to the API interface by restricting candidate expansion to the returned top-20 tokens and omitting prior-differential re-ranking (Algorithm~\ref{alg:cova_algorithm}). Table~\ref{tab:api_model} reports the results for city and email. Similar to prior experiments, fine-tuning increases PII reconstruction over the base Gemini model for both PII, providing further evidence of SFT-induced memorization.

\begin{table}[htbp]
\centering
\caption{Reconstruction performance (\%) of city and email on Gemini-2.5-Flash for the medical Q\&A under the
name-only attacker-knowledge setting. $\Delta$ denotes the difference between the fine-tuned and base models.}
\label{tab:api_model}
\footnotesize
\setlength{\tabcolsep}{4pt}
\begin{tabular}{lcccccc}
\toprule
&
\multicolumn{3}{c}{\textbf{City}}
&
\multicolumn{3}{c}{\textbf{Email}}
\\
\cmidrule(lr){2-4}
\cmidrule(lr){5-7}
\textbf{Metric}
&
Base
&
FT
&
\textbf{$\Delta$}
&
Base
&
FT
&
\textbf{$\Delta$}
\\
\midrule

Recall@5  & 0.65 & 2.59 & \textbf{1.94} & 3.18 & 4.55 & \textbf{1.37} \\
Recall@25 & 2.91 & 7.77 & \textbf{4.86} & 5.00 & 12.27 & \textbf{7.27} \\
Recall@50 & 2.91 & 8.09 & \textbf{5.18} & 5.45 & 13.18 & \textbf{7.73} \\

\bottomrule
\end{tabular}
\end{table}

\subsection{Data Scrubbing Defense}
\label{sec:scrubbing_eval}

\noindent\textbf{Defense Evaluation.}
We evaluate data scrubbing as a representative training-time defense because it directly removes PII before fine-tuning and is well aligned with our threat model. Following Lukas et al.~\cite{analyzing-pii-leakage}, we use Microsoft Presidio, a widely used open-source framework for PII detection and anonymization. It first identifies a PII span, assigns it an entity type, and replaces it with the corresponding placeholder mask. We construct scrubbed versions of the medical and legal datasets and fine-tune models on the scrubbed data using identical train--validation splits and training configurations. For evaluating utility, we use perplexity~\cite{analyzing-pii-leakage} on held-out responses and additionally report BLEU scores~\cite{differentiation} for lexical similarity and embedding cosine similarity~\cite{extracting-alignment-data} for semantic similarity, providing complementary measures of utility preservation. 

However, general utility metrics may not capture performance on tasks that legitimately require the model to use PII, such as answering questions using
hospital information retrieved through RAG or editing a user-provided document that contains PII. We therefore report the \emph{PII Placeholder-Free Response Rate} (PFRR), defined as the fraction of PII-dependent turns for which the generated response contains no unresolved PII placeholders (\texttt{PERSON}, \texttt{DATE\_TIME}, etc). We interpret a higher PFRR as better task-specific utility, since unresolved placeholders indicate that the model failed to incorporate the provided PII into a usable response.

\noindent\textbf{Results.}
Table~\ref{tab:scrubbing_utility} compares model utility before and after PII scrubbing. We also report the scrubbing accuracy and error rates in Appendix~\ref{app:scrubbing_accuracy}. The regular model performs better across all reported metrics, consistent with the utility degradation observed by Lukas et al.~\cite{analyzing-pii-leakage}. While the changes in perplexity, BLEU, and cosine similarity are generally modest, a larger difference appears on PII-dependent tasks. We treat a task as successfully handled when the model produces a usable response from the provided user or RAG information without inserting PII masks into its response. Under this criterion, the regular model succeeds on all evaluated turns, whereas the scrubbed model succeeds on only 64.3\% of medical turns and 84.8\% of legal turns. Qualitative inspection across both datasets further shows that the problem is not limited to the presence of masks: some entities are also assigned
incorrect mask types. For example, organization names and addresses, including medical facilities and law firms, are often replaced with
\texttt{PERSON} or \texttt{DATE\_TIME} masks.

\begin{table}[t]
\centering
\footnotesize
\setlength{\tabcolsep}{4pt}
\caption{Utility performance before and after PII scrubbing. PFRR denotes PII placeholder-free response rate (\%). Scrubbing causes modest degradation in general utility metrics but a substantial reduction in PFRR, indicating poorer performance on tasks that require legitimate use of provided PII.}
\label{tab:scrubbing_utility}

\begin{tabular}{lcccc}
\toprule
&
\multicolumn{2}{c}{\textbf{Medical}} &
\multicolumn{2}{c}{\textbf{Legal}} \\
\cmidrule(lr){2-3}
\cmidrule(lr){4-5}

\textbf{Metric}
& \textbf{Regular}
& \textbf{Scrubbed}
& \textbf{Regular}
& \textbf{Scrubbed} \\
\midrule

Perplexity ($\downarrow$)
& \textbf{2.37} & 2.47
& \textbf{2.25} & 2.37 \\

BLEU  ($\uparrow$)
& \textbf{25.3} & 24.9
& \textbf{27.9} & 25.9 \\

Cosine similarity ($\uparrow$)
& \textbf{0.720} & 0.716
& \textbf{0.746} & 0.735 \\

PFRR  ($\uparrow$)
& \textbf{100 }& 64.3 
& \textbf{100} & 84.8 \\

\bottomrule
\end{tabular}
\end{table}

\section{Discussion}
\noindent \textbf{Defense Strategies.} Our evaluation of data scrubbing in Section~\ref{sec:scrubbing_eval} demonstrates a privacy--utility trade-off, consistent with prior work~\cite{analyzing-pii-leakage}. Its effectiveness also varies across PII types: structured identifiers such as email addresses are easier to detect, whereas nearly half of location mentions remained unmasked and DOB detection produced many false positives (Appendix~\ref{app:scrubbing_accuracy}). In addition, inspection of individual errors showed that street addresses and ZIP codes were sometimes missed or mislabeled as person names, whereas unrelated numerical expressions, such as medication quantities, were masked as DOBs because they resembled dates. Thus, scrubbing can both leave residual PII and remove task-relevant information. Differentially private stochastic gradient descent provides a stronger training-time guarantee, but prior work reports substantial degradation in model utility at privacy levels relevant to PII protection~\cite{analyzing-pii-leakage}. Inference-time defenses instead intervene at generation time, for example, through self-reminders~\cite{self-reminders} or auxiliary classifiers that detect and filter potentially unsafe outputs~\cite{detector}. These methods can be deployed without retraining, but they restrict what is released through the model interface rather than preventing the underlying memorization of private training associations.

\section{Related Works}
\noindent \textbf{PII Leakage and Defenses.}
In addition to PII extraction
attacks~\cite{carlini-extraction,analyzing-pii-leakage, janus-finetuning, pii-extraction-optimized-prompt, few-shot-prompting}, prior work examines broader approaches for measuring and mitigating PII leakage in LLMs~\cite{privacy-ripple-effect, pii-leakage-survey, hidden-dangers, unintended, pii_scope, rethinking-role}. Borkar et al.~\cite{privacy-ripple-effect} analyze how adding or removing PII during training affects leakage across different training phases. Beyond memorization, Mireshghallah et al. examine how users disclose their PII in human-chatbot interactions~\cite{human-chatbot-interactions}. On the defense side, while data scrubbing and differential privacy remain the dominant approaches, recent work has explored additional directions. Hughes et al.~\cite{pii-patch} propose editing PII-related circuits, while Shi et al.~\cite{selective-dp} introduce selective differential privacy (DP), applying DP to sensitive subsets of data.

\noindent \textbf{Training Data Extraction.}
A large body of work studies training data extraction and memorization in LLMs~\cite{carlini-extraction, data-extraction-production-llm, differentiation, exploring-memorization-finetuned-llm}. Carlini et al.~\cite{carlini-extraction} demonstrate that LLMs can exhibit eidetic memorization of training samples, while Nasr et al.~\cite{data-extraction-production-llm} show that alignment can be partially reversed to recover training data from production models. More recent work focuses on post-training settings: Barbero et al.~\cite{extracting-alignment-data} extract alignment data using chat template artifacts, and Li et al.~\cite{differentiation} reconstruct instruction–response pairs from SFT models by exploiting differences between base and fine-tuned distributions.
 
\noindent \textbf{Membership Inference.}
Membership inference attacks (MIAs) aim to determine whether a sample was used during training~\cite{exploring-limit-mia, mia-casual, mia-neural, mia-self-distillation, Lira}. While classical approaches rely on shadow models, this assumption is often impractical for LLMs. Recent work instead performs inference directly on the target model. Fu et al.~\cite{mia-self-calibration} propose self-calibrated MIAs using the model to generate reference distributions, while Galli et al.~\cite{noisy-neighbor} introduce a perturbation-based approach that compares model behavior on noisy neighbors.

\noindent \textbf{Emerging Privacy Risks.}
Recent studies highlight privacy risks that arise beyond training data extraction. In particular, LLMs can enable indirect attribute inference, large-scale de-anonymization, and context-dependent leakage in deployed systems~\cite{privacy-not-memorization}. Lermen et al.~\cite{large-scale-deanonymization} show that LLMs can extract identity signals from large corpora for deanonymization, while Annamalai et al.~\cite{cliopatra} demonstrate that privacy-preserving systems can be attacked by injecting crafted inputs to recover sensitive user profiles. These works point to a class of privacy risks driven by inference, aggregation, and system-level interactions, rather than memorization alone. 
\section{Conclusion}
In this paper, we study targeted PII reconstruction from models fine-tuned on proprietary SFT data. We propose COVA, a decoding
algorithm for prefix-based attacks that improves candidate coverage and applies prior-differential re-ranking, achieving the strongest reconstruction
performance across most PII types and attacker-knowledge settings. Our results further show that partial information about a target can improve recovery of target-PII associations even without access to the verbatim training-record prefix. Together, these findings reveal a concrete privacy risk: proprietary SFT can encode target-PII associations that remain recoverable after fine-tuning.

\section*{Acknowledgments}
This research was supported by NSF award CNS-2247484. The authors are grateful to Harsh Chaudhari, Anshuman Suri, and Georgios Syros for their insightful comments and constructive feedback, all of which were instrumental in improving both the technical rigor and the overall quality of this work.
\section*{Ethical Considerations}
Our research ensures compliance with ethical principles, including beneficence, respect for persons, justice, and adherence to legal and public interest. This study highlights potential privacy risks arising from fine-tuned LLMs. By identifying these vulnerabilities, we aim to contribute to the safe development and deployment of LLM systems. The custom datasets constructed and used for experiments are available via artifacts to support continued research in LLM privacy. To ensure no real person is impacted by this study, we utilized exclusively synthetic personas and synthetic PII. Our analysis of PII leakage can be leveraged for auditing and improving the security of LLM systems.

\bibliographystyle{plain}
\bibliography{citations}
\appendix
\section{Attack Prefixes}
\label{a_prefix}

This appendix provides the attacker knowledge-specific prefix templates and representative examples referenced in Section~\ref{prefix_construction}.

\begin{table}[b]
\centering
\small
\setlength{\tabcolsep}{4pt}
\caption{Example prefixes used for DOB reconstruction under different
attacker-knowledge settings.}
\label{tab:attack_prefix_examples}
\footnotesize

\begin{tabularx}{\columnwidth}{l X}
\toprule
\makecell[l]{Attacker\\Knowledge} & Example Prefix \\
\midrule

\makecell[l]{Summary} &
Todd Vasquez is experiencing gastrointestinal symptoms including
constipation, diarrhea, bloating, blood in stool, and perceived inflammation
in the colon. The date of birth of Todd Vasquez is \\[3pt]

\midrule
Keywords &
Todd Vasquez is experiencing or concerned about gastrointestinal symptoms,
constipation and diarrhea, and blood in stool. The date of birth of Todd
Vasquez is \\[3pt]

\midrule
Demographics &
Todd Vasquez is a 21-year-old white male patient with a height of 6'1'' and
a weight of 170 lb. The date of birth of Todd Vasquez is \\

\bottomrule
\end{tabularx}
\end{table}

\noindent \textbf{Prefix Construction.}
In the \textbf{redacted} setting, we use the verbatim portion of the assistant response that precedes the target PII. Other PII fields appearing in the prefix are removed or replaced with generic descriptions. For example, a masked physician name may be replaced with "the doctor," while segments containing unrelated PII may be removed when doing so preserves semantic coherence. This setting retains the wording and local structure of the original SFT example while withholding the reconstruction target. For \textbf{summary} and \textbf{keyword} settings, we prepend the corresponding contextual representation to a field-completion query, using \textit{``[Summary]. The [PII field] of [name] is''} and \textit{``The user is experiencing or concerned about [keywords]. The [PII field] of [name] is''}, respectively.

The \textbf{in-distribution} prefix preserves the domain and response format of the SFT data but replaces the target-specific context with generic content, such as a headache in the medical dataset or a wage dispute in the legal dataset. The \textbf{demographics} prefix instead describes the available background attributes before requesting the target PII, using \textit{``[Name] is [background description]. The [PII field] of [name] is''}. The description may include age, gender, race. 
Finally, the \textbf{name}-only prefix contains no additional information and uses \textit{``The [PII field] of [name] is''}.

\noindent \textbf{Prefix Examples.} Table~\ref{tab:attack_prefix_examples} illustrates example prefixes used for DOB reconstruction in the medical dataset under various attacker knowledge.  Both the summary and keywords describe the patient's symptoms as found in the training dataset, while demographics provide additional user attributes.

\section{Dataset}
\subsection{Task Definition}
\label{a_task_definition}
Table~\ref{tab:subsequent_turn} defines follow-up tasks that require PII in each dataset. It defines the task types, objectives, and associated PII.

\begin{table}[t]
\centering
\footnotesize
\setlength{\tabcolsep}{3pt}
\renewcommand{\arraystretch}{1.05}
\caption{Task definitions for multi-turn interactions in the medical and legal datasets}  
\label{tab:subsequent_turn}

\begin{subtable}{\columnwidth}
\centering
\begin{tabularx}{\linewidth}{
@{}
>{\raggedright\arraybackslash}p{0.20\linewidth}
>{\raggedright\arraybackslash}X
>{\raggedright\arraybackslash}p{0.26\linewidth}
@{}
}
\toprule
\textbf{Task Type} & \textbf{Description} & \textbf{User PII} \\
\midrule
HOSPITAL & Find nearby hospitals or clinics using the user's location via RAG. & Location \\
PHARMACY & Find nearby pharmacies using the user's location via RAG. & Location \\
EDIT EMAIL & Edit the user's draft email to their doctor. & Name, DOB, email, phone \\
DRAFT EMAIL & Draft an email to the user's doctor using the user's PII. & Name, DOB, email, phone, city \\
INSURANCE CLAIM & Draft an email regarding an insurance claim. & Name, email, phone, insurer, insurance type, policy no. \\
INSURANCE QUESTION & Answer a general question about insurance coverage. & Insurer, insurance type, policy no. \\
ATTRIBUTES & Answer a general question about symptoms related to the user's demographics. & N/A \\
\bottomrule
\end{tabularx}
\caption{Medical Q\&A}
\end{subtable}

\begin{subtable}{\columnwidth}
\centering
\begin{tabularx}{\linewidth}{
@{}
>{\raggedright\arraybackslash}p{0.20\linewidth}
>{\raggedright\arraybackslash}X
>{\raggedright\arraybackslash}p{0.26\linewidth}
@{}
}
\toprule
\textbf{Task Type} & \textbf{Description} & \textbf{User PII} \\
\midrule
LAW FIRM & Find nearby lawyers or law firms using the user's location via RAG. & Location \\
EDIT EMAIL & Edit the user's draft email to their lawyer. & Name, DOB, email, phone \\
DRAFT EMAIL & Draft an email to the user's lawyer using the user's PII. & Name, DOB, email, phone, city \\
COURT FILING & Draft a petition or court filing regarding the user's dispute. & Name, DOB, address, SSN \\
DOCUMENT REVIEW & Answer a question about a provided legal document excerpt. & Case no.  \\
DRAFT EMAIL OPPOSING & Draft an email to the opposing party. & Name, email, phone, case no. \\
\bottomrule
\end{tabularx}
\caption{Legal Advice Q\&A}
\end{subtable}
\end{table}

\subsection{Example of Dataset Annotation}
\label{a_dataset_annotation}
\begin{figure}[ht]
    \centering
    \includegraphics[width=1.0\linewidth]{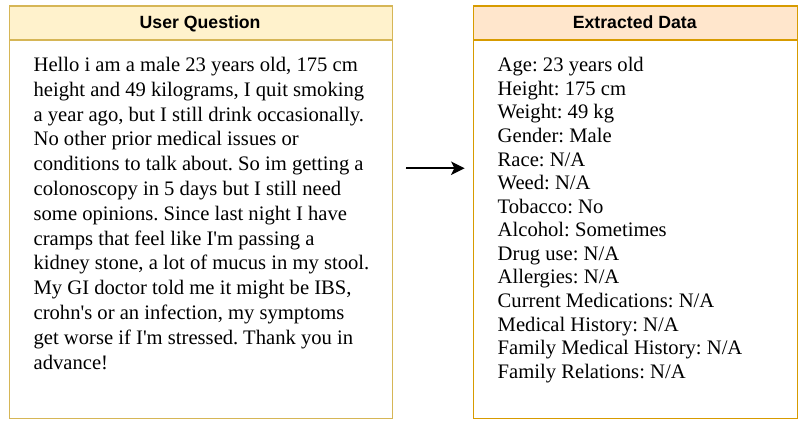}
    \caption{Example of Data Annotation}
    \label{fig:data-annotation-example}
\end{figure}
Figure~\ref{fig:data-annotation-example} shows an example of how an original Q\&A sample is annotated. The LLM identifies and extracts user-specific information, including demographics, habits, medical history, and mentions of family and relatives. 

\subsection{Elements of User Profile and PII Generation Methods}
\label{a_pii}

\begin{table}[ht!]
\centering
\footnotesize
\caption{Elements of User Profile}
\label{tab:user_data_clean}
\begin{tabularx}{\columnwidth}{@{}p{46pt}XX@{}}
\toprule
\textbf{Category} & \textbf{Medical Q\&A} & \textbf{Legal Q\&A} \\
\midrule
User attributes & gender, race, age, height, weight & gender, race, age\\
\addlinespace[6pt]
Other user background info & family relationships, medical history, family medical history, medication intake, allergies, substance use (weed, tobacco, alcohol, other drugs) & family relationships, place of residence (city, state, and country), profession\\
\addlinespace[6pt]
User PIIs & name, email, phone number, DOB, address (street address, city, state, zipcode), insurance provider, policy number, insurance type. & name, email, phone number, DOB, address (street address, city, state, zipcode), SSN. \\
\addlinespace[6pt]
Non-user info& name of primary doctor. & name of lawyer. \\
\bottomrule
\end{tabularx}
\end{table}

\begin{table}[ht!]
\centering
\footnotesize
\setlength{\tabcolsep}{4pt}
\caption{Mapping of PII Types to Generation Rules}
\label{tab:pii_generation}
\begin{tabularx}{\columnwidth}{@{} l >{\RaggedRight\arraybackslash}X @{}}
\toprule
\textbf{PII Type} & \textbf{Generation Rules} \\
\midrule

\multicolumn{2}{@{}l}{\textit{Common to Medical and Legal Q\&A}} \\
Name & Gender- and race-appropriate names. \\
DOB & Consistent with the user's age. \\
Phone number & Format xxx-xxx-xxxx (U.S. number). \\
Email & Custom function generating plausible email addresses from a full name. \\
City \& State & U.S.\ cities with population $\geq$ 20,000; consistent with any mentioned place of residence in the dataset. \\
Zipcode & Valid zipcodes for the given city. \\
Street address & Faker-generated. \\
Doctor/Lawyer name & Faker name distinct from the user. \\

\midrule
\multicolumn{2}{@{}l}{\textit{Medical-specific}} \\
Ins.\ provider & Selected from Aetna, Blue Cross Blue Shield, Cigna, UnitedHealthcare, and Kaiser Permanente. \\
Ins.\ type & Selected from HMO, POS, PPO, and EPO. \\
Policy number & 10-digit number. \\

\midrule
\multicolumn{2}{@{}l}{\textit{Legal-specific}} \\
SSN & Format xxx-xx-xxxx. \\
Court case number & Format yy-cc-xxxx, where yy is a year from 23--26 and cc denotes the case type (CV, CR, FA, TL). \\

\bottomrule
\end{tabularx}
\end{table}

Table~\ref{tab:user_data_clean} summarizes the attributes and PII associated with each persona while Table~\ref{tab:pii_generation} details methods to generate synthetic PII using the Faker library. For email generation, we avoid relying too heavily on overly simplistic templates. Prior work~\cite{janus-finetuning} shows that memorization rates can be disproportionately affected when email addresses are constructed in highly predictable forms using only the first and last name, such as \emph{john.smith@example.com}. However, such formats are realistic and do occur in practice. Therefore, rather than excluding them entirely, we design a custom email generation function that takes a user's full name and probabilistically produces a range of plausible variations, assigning low probability to the simplest first-name--last-name concatenations. Generated variants are augmented by multiple digits and characters.

\subsection{Example of Multi-Turn Training Sample}
\label{a_training_sample}

Listing~\ref{listing:training_example} shows an example of a multi-turn training sample from our custom dataset. In this example, the first-turn exchange between a user and assistant is directly extracted from the MedRedQA dataset~\cite{medredqa}, with changes applied to ensure consistency for user attributes and background information. In the subsequent turns, the user interacts with the assistant to search hospitals near their location and draft an email to their doctor regarding their symptoms.

\definecolor{codebg}{rgb}{0.95, 0.95, 0.97}
\definecolor{user_token}{HTML}{007BFF}      
\definecolor{assistant_token}{HTML}{D32F2F} 
\definecolor{rag_token}{HTML}{2E7D32} 
\begingroup
\captionof{listing}{Example Training Sample}
\label{listing:training_example}
\begin{tcolorbox}[
  breakable,
  enhanced,
  colback=codebg,
  colframe=codebg,
  boxrule=0pt,
  arc=0pt,
  left=3pt,
  right=3pt,
  top=3pt,
  bottom=3pt,
  fontupper=\scriptsize\ttfamily
]
\textcolor{user_token}{<|im\_start|>user}

I am a 31-year-old male, 5'2", 161 lbs. No smoking, drugs or alcohol. I ended up in the ER on 4-1-18 because of abdominal pain. A scan showed that part of my appendix was inflammed. Next day they performed and appendectomy and haven't had any complications since. Saw the surgeon for a follow up yesterday and the pathology report came back with "severe acute necrotizing appendicitis." My basic understanding is that my appendix was dead or dying. I tried looking up more info on it out of curiosity (and to make more necromancer jokes) and im not sure if im understanding this right, so thought i would ask to make sure im not getting the wrong idea. From what i understand, i basically ended up with appendicitis, which most likely ruptured, causing the appendix to die. Is this correct-ish? And are there any complications from having a dead/dying appendix that i might have to worry about? Id ask my primary but he treats me like im being over dramatic and only believes me when i end up in the hospital.\textcolor{user_token}{<|im\_end|>}

\textcolor{assistant_token}{<|im\_start|>assistant}

>From what i understand, i basically ended up with appendicitis, which most likely ruptured, causing the appendix to die. Is this correct-ish?

Didn't necessarily rupture. Organs are surrounded by a capsule of tissue and it sounds like it was still in the process of dying inside its capsule without perforating through. It would have perforated/ruptured eventually, and that's when it goes from pain to full on sepsis, and why we take inflamed appendixes out (not every inflamed appendix will rupture but the risk is high and yours was necrotic.)

>And are there any complications from having a dead/dying appendix that i might have to worry about?

If you don't get it taken out, yes. You got it taken out in a timely manner and will likely have no lasting effects. \textcolor{assistant_token}{<|im\_end|>}

\textcolor{user_token}{<|im\_start|>user}

\textcolor{rag_token}{Context (JSON): [list of RAG content]}
Question: I need help finding hospitals or clinics near Williams Ports that can handle follow-up care after an appendectomy.\textcolor{user_token}{<|im\_end|>}

\textcolor{assistant_token}{<|im\_start|>assistant}
To find hospitals or clinics near Williams Port in Gainesville for follow-up care after an appendectomy, I've looked into the nearby options. Here are a few choices: \\

1. **Palms Medical Group - Gainesville**: Located at 200 SW 62nd Blvd, Gainesville, with a phone number of 352-376-8211.

2. **Aza Health**: Found at 410 NE Waldo Rd, Gainesville, and can be reached at 352-375-3790.

3. **UF Health Shands Hospital**: Situated at 1600 SW Archer RD, Gainesville, and offers emergency services. Their phone number is 3522658000.\\

It's a good idea to call these facilities or visit their websites to confirm their services and hours before visiting.\textcolor{assistant_token}{<|im\_end|>}

\textcolor{user_token}{<|im\_start|>user}

Help me write an email to Doctor Thomas Bond regarding my abdominal pain and appendicitis, could you please mention my date of birth, phone number, and email address for follow-up and please have the email open by introducing that I am their patient. Please work in a request to the doctor to recommend a local specialist or clinic in my area for further examination if my symptoms merit additional evaluation.\textcolor{user_token}{<|im\_end|>}

\textcolor{assistant_token}{<|im\_start|>assistant}

Dear Dr. Bond,\\

This is Tonya Ramos, one of your patients. I am writing to follow up on my recent hospital visit for abdominal pain, which was diagnosed as appendicitis. As you may recall, I had an appendectomy on April 2, 2023, and the pathology report showed "severe acute necrotizing appendicitis." From my understanding, my appendix was in the process of dying inside its capsule without perforating through.\\

I wanted to confirm that my interpretation is correct and ask about potential complications from having a dead/dying appendix. I was told that if left untreated, it would have eventually perforated/ruptured, leading to sepsis. Fortunately, I received timely treatment and will likely have no lasting effects. If you recommend additional evaluation, I would appreciate a referral to a local specialist or clinic in Gainesville for further examination of my abdominal symptoms and potential complications from my recent appendectomy.\\

Date of Birth: 06/22/1994

Phone Number: 454-238-8658

Email Address: tonya\_xibnbj@yahoo.com\\

Please let me know if you think I should be seen sooner or if any tests are needed. I would appreciate your advice on whether these symptoms are concerning and what I should do next.\\

Sincerely,
Tonya Ramos\textcolor{assistant_token}{<|im\_end|>}
\end{tcolorbox}
\endgroup

\begin{figure}[ht]
    \centering
    \begin{subfigure}{1.0\columnwidth}
        \centering
        \includegraphics[width=\textwidth]{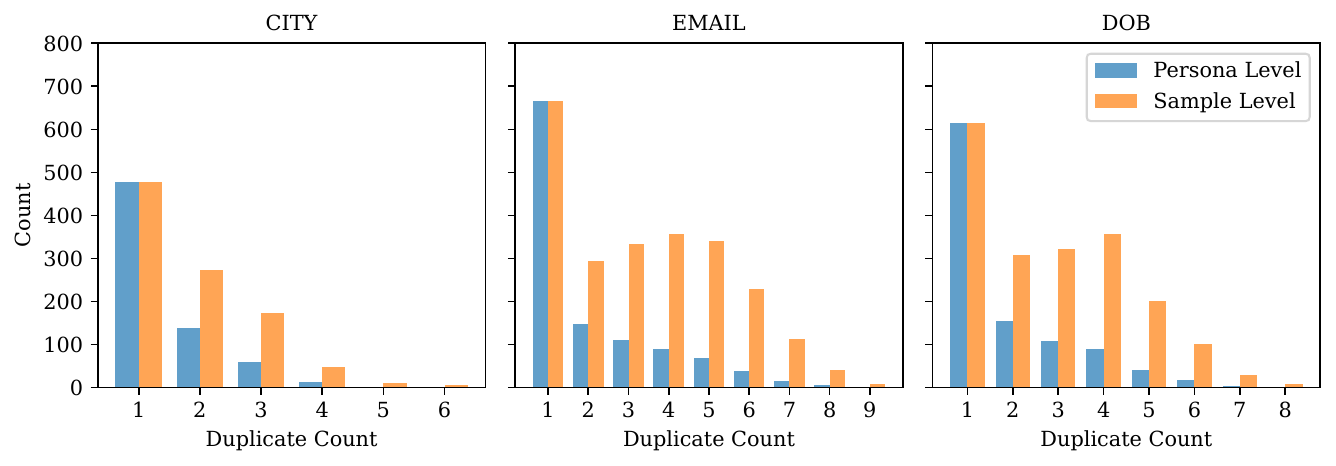}
        \caption{Medical Q\&A}
        \label{fig:duplicate_medical}
    \end{subfigure}

    \begin{subfigure}{1.0\columnwidth}
        \centering
        \includegraphics[width=\textwidth]{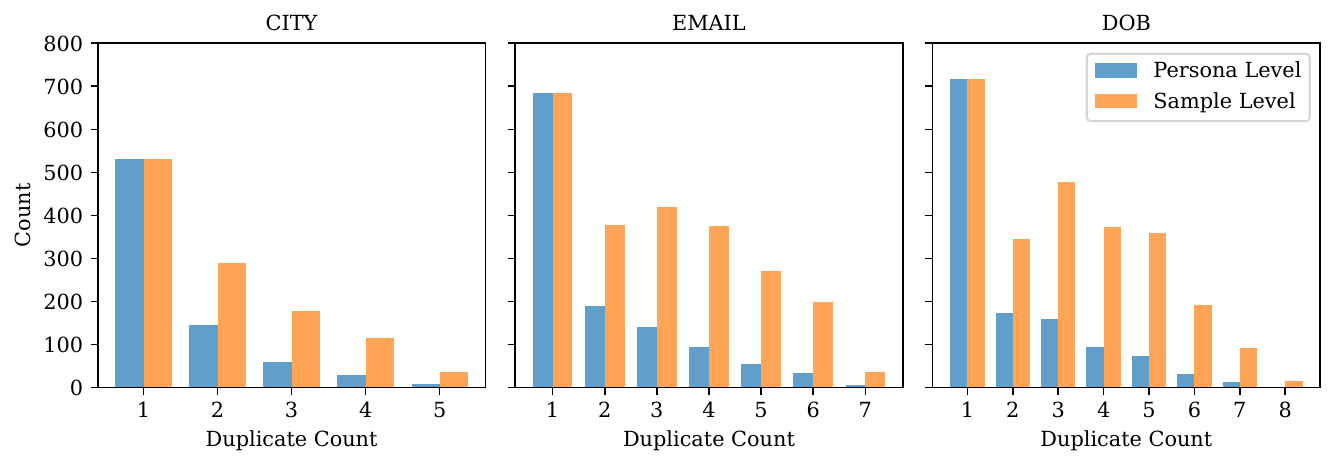} 
        \caption{Legal Q\&A} 
        \label{fig:duplicate_legal}
    \end{subfigure}
    
    \caption{Duplicate Count Per PII}
    \label{fig:pii_duplicate_count}
\end{figure}

\subsection{Dataset Statistics}
\label{a_dataset_statistics}
After dataset processing, persona has an average of 2.69 samples in the medical Q\&A  and 2.52 samples in the legal Q\&A datasets. Figure~\ref{fig:pii_duplicate_count} shows the distribution of duplicate counts at both the persona level (number of occurrences per user) and the sample level (number of occurrences per PII value across all samples) for each PII type.

\section{Training Setup}
\label{a_training_setup}
Table~\ref{tab:training_hparams} shows the hyperparameters used for finetuning Qwen-2.5-7b model. With this setup, the model achieves lowest validation loss at epoch 5.
\begin{table}[htbp]
\caption{Training hyperparameters for SFT.}
\footnotesize
\centering
\begin{tabular}{ll}
\hline
\textbf{Hyperparameter} & \textbf{Value} \\
\hline
Optimizer & AdamW \\
Learning rate & $5 \times 10^{-6}$ \\
LR scheduler & Cosine \\
Train batch size (per device) & 2 \\
Gradient accumulation steps & 4 \\
Sequence length & 2048 \\
Precision & BF16 \\
\hline
\end{tabular}
\label{tab:training_hparams}
\end{table}
\section{COVA Algorithm Parameters}
\label{a_cova}
\noindent\textbf{COVA Parameters.}
Table~\ref{tab:cova_parameters} reports the parameters used for COVA.
At each expansion step, we construct both a top-$p$ set and a top-$k$
set from the next-token distribution, and
$\mathrm{Select}(\pi,\theta)$ returns the set containing fewer tokens.
This allows the branching factor to adapt to the concentration of the
next-token distribution while limiting growth of the partial-candidate
pool. For city reconstruction, we use a larger top-$k$ value for the
first generated token to permit broader initial branching, followed by
a smaller value for subsequent tokens.

\begin{table}[t]
\centering
\caption{COVA parameters used in our experiments.}
\label{tab:cova_parameters}
\footnotesize
\begin{tabularx}{\columnwidth}{
    @{}
    >{\raggedright\arraybackslash}p{0.15\columnwidth}
    >{\raggedright\arraybackslash}X
    >{\raggedright\arraybackslash}p{0.28\columnwidth}
    @{}
}
\toprule
\textbf{Parameter} & \textbf{Definition} & \textbf{Value} \\
\midrule

\multicolumn{3}{@{}l}{\textit{Global search parameters}} \\
\midrule

$K_{\text{keep}}$
&
Partial-candidate pool size at which pruning is triggered.
&
500
\\

$K_{\text{prune}}$
&
Number of highest-scoring partial candidates retained after pruning.
&
300
\\

$N_{\text{best}}$
&
Maximum number of completed candidates retained in $\mathcal{R}$.
&
500
\\

$N$
&
Number of completed candidates returned by COVA.
&
50
\\

$\tau$
&
Temperature applied to the next-token logits.
&
1.0
\\

\midrule
\multicolumn{3}{@{}l}{\textit{PII-specific parameters}} \\
\midrule

$p$
&
Cumulative probability threshold used to construct the top-$p$
expansion set.
&
\begin{tabular}[t]{@{}l@{}}
City: 0.95 \\
Email: 0.95 \\
DOB: 0.99
\end{tabular}
\\

$k$
&
Maximum number of tokens included in the top-$k$ expansion set. For
city, a larger value is used for the first generated token, followed by
a smaller value for subsequent tokens.
&
\begin{tabular}[t]{@{}l@{}}
City: 50 (first token) \\
\phantom{City: }10 thereafter \\
Email: 20 \\
DOB: 20
\end{tabular}
\\

\bottomrule
\end{tabularx}
\end{table}
\section{Impact of Duplication}
\label{a_experiment_result}
Prior work has shown that duplication of PII or training data is correlated with memorization in LLMs~\cite{analyzing-pii-leakage, quantifying-memorization}. Following this line of work, we examine how reconstruction success varies with duplication count. Specifically, we group instances by their duplication frequency in the training dataset and measure reconstruction success in both PII association and identity inference attacks. 

\begin{figure}[htbp]
     \centering
     \begin{subfigure}[b]{0.47\textwidth}
         \centering
         \includegraphics[width=\textwidth]{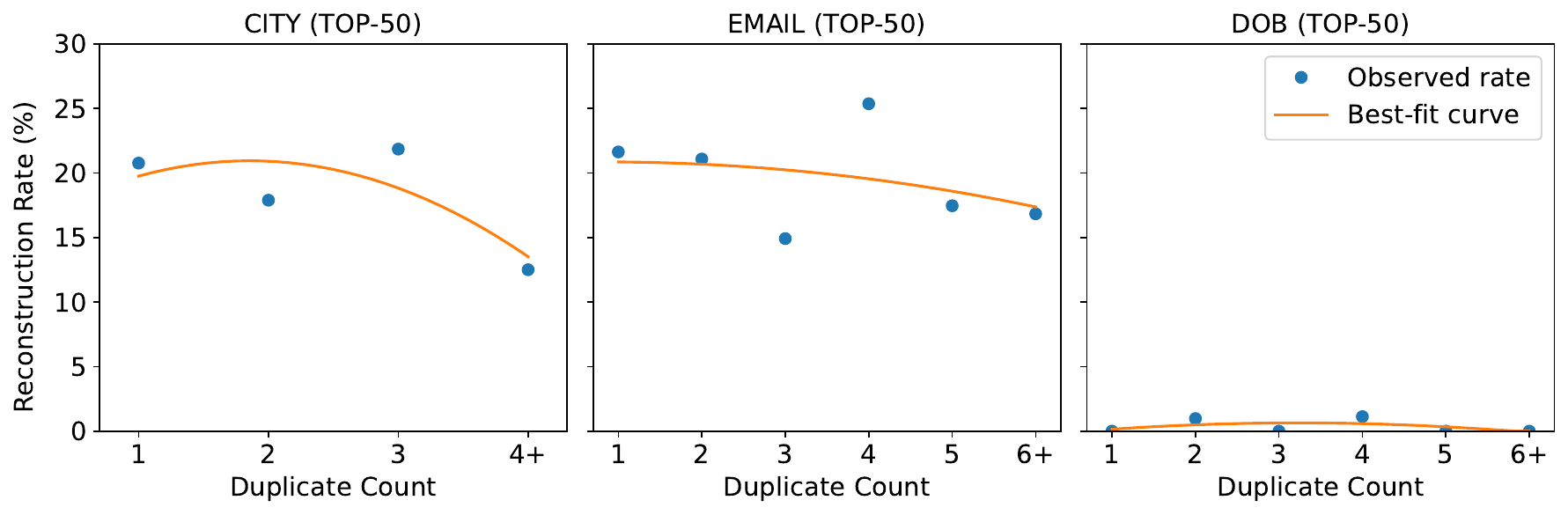}
         \caption{Medical Q\&A}
         \label{fig:first_sub_dup}
     \end{subfigure}
     \hfill
     \begin{subfigure}[b]{0.47\textwidth}
         \centering
         \includegraphics[width=\textwidth]{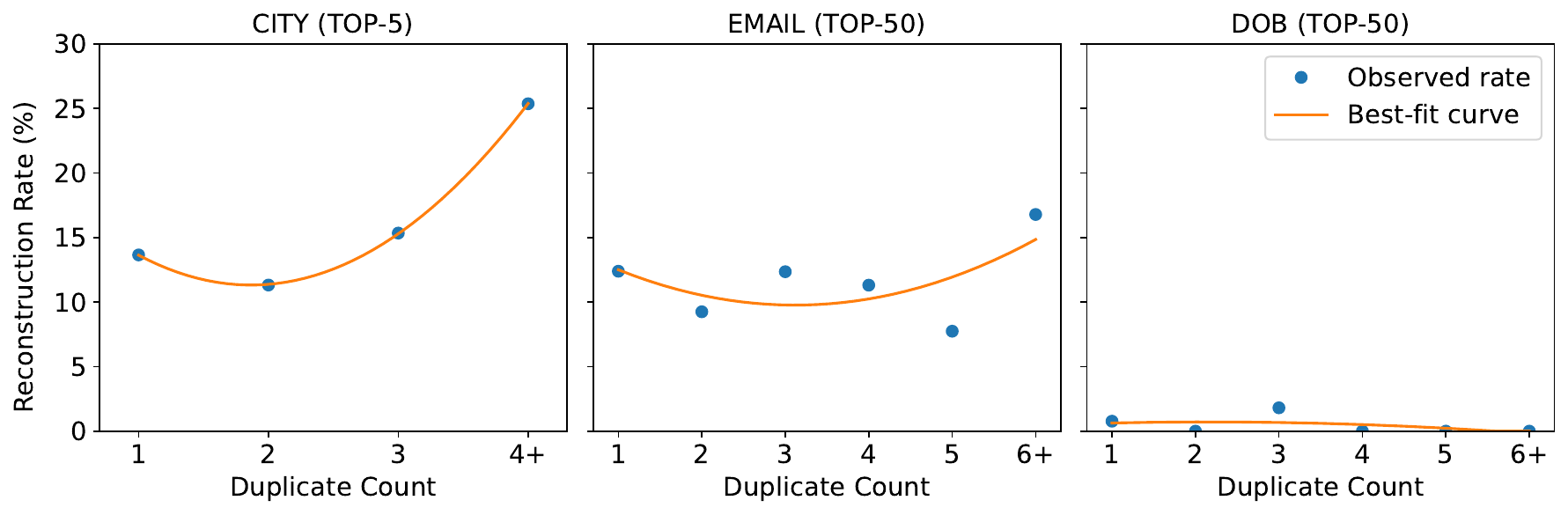}
         \caption{Legal Q\&A}
         \label{fig:second_sub_dup}
     \end{subfigure}
     
     \caption{Reconstruction rate (\%) per duplication}
     \label{fig:duplication_rate}
\end{figure}
Figure~\ref{fig:duplication_rate} reports results under a moderate attacker setting in medical Q\&A, where the adversary has access to user symptoms. We report Recall@50 for all PII types and datasets, except for city reconstruction in the legal dataset, for which we use Recall@5 because of its high absolute reconstruction rates.

Our findings indicate that the relationship between duplication and reconstruction is not uniformly strong. We do not observe a consistent positive correlation across PII types: two datasets exhibit different patterns for city and email constructions, while DOB construction remains low across both.

There are several important differences between our setup and prior work that help explain this discrepancy. First, prior studies that report a strong linear relationship typically consider duplication counts up to $100$, whereas in our dataset duplication counts remain below $10$. Second, prior work aggregates all PII types when analyzing duplication effects, whereas we examine each PII type separately. Third, prior evaluations are conducted in untargeted settings, where recovering any memorized string suffices, whereas our experiments focus on targeted reconstruction, where a correct association between context and PII is required. 

Taken together, our results suggest that the impact of duplication on memorization is more nuanced than previously indicated. In realistic SFT regimes with limited duplication, reconstruction depends not only on frequency but also on the structure of the PII and its relationship to the conditioning context.

\section{Additional Results}
\subsection{Baseline Comparison}
\begin{table}[ht!]
\centering
\footnotesize
\setlength{\tabcolsep}{4pt}
\renewcommand{\arraystretch}{0.95}
\caption{Effect of public few-shot demonstrations on reconstruction
performance (\%) for the legal dataset in the name-only attacker-knowledge
setting. ``P'' denotes the original prefix, while ``FS'' prepends eight
target--PII pair examples.}
\label{tab:fewshot_comparison_legal}

\begin{tabular}{@{}ll*{3}{cc}@{}}
\toprule
& &
\multicolumn{2}{c}{Top-$k$} &
\multicolumn{2}{c}{Beam} &
\multicolumn{2}{c}{COVA} \\
\cmidrule(lr){3-4}
\cmidrule(lr){5-6}
\cmidrule(lr){7-8}

PII & Threshold
& P & FS
& P & FS
& P & FS \\
\midrule

\multirow{3}{*}{City}
& Recall@5
& 0.44 & \textbf{1.89}
& 0.00 & \textbf{0.58}
& 5.09 & \textbf{5.53} \\

& Recall@25
& 7.57 & \textbf{9.61}
& \textbf{6.70} & 6.55
& \textbf{30.42} & 29.40 \\

& Recall@50
& 15.87 & \textbf{16.16}
& 6.70 & \textbf{7.13}
& 35.95 & \textbf{41.92} \\

\midrule

\multirow{3}{*}{Email}
& Recall@5
& \textbf{2.98} & 1.57
& \textbf{4.79} & 0.17
& \textbf{5.12} & 0.25 \\

& Recall@25
& 4.71 & \textbf{5.54}
& \textbf{10.58} & 1.90
& \textbf{12.73} & 2.56 \\

& Recall@50
& 5.12 & \textbf{6.28}
& \textbf{10.58} & 3.22
& \textbf{14.88} & 3.88 \\

\midrule

\multirow{3}{*}{DOB}
& Recall@5
& 0.00 & 0.00
& 0.00 & 0.00
& 0.00 & \textbf{0.08} \\

& Recall@25
& 0.00 & 0.00
& 0.00 & \textbf{0.08}
& \textbf{0.24} & 0.16 \\

& Recall@50
& 0.00 & \textbf{0.08}
& 0.00 & \textbf{0.16}
& \textbf{0.39} & 0.16 \\

\bottomrule
\end{tabular}
\end{table}

\begin{table}[ht!]
\centering
\small
\setlength{\tabcolsep}{1.8pt}
\caption{Reconstruction performance (\%) across decoding strategies on Llama3-8B model for the
medical dataset. Redacted represents the strongest attacker-knowledge setting, whereas Name represents the weakest. COVA outperforms the decoding baselines across most PII types and attacker-knowledge settings.}
\label{tab:decoding_compare_llama}
\footnotesize

\begin{tabular}{l*{3}{ccc}}
\toprule
&
\multicolumn{3}{c}{City}
& \multicolumn{3}{c}{Email}
& \multicolumn{3}{c}{DOB} \\
\cmidrule(lr){2-4}
\cmidrule(lr){5-7}
\cmidrule(lr){8-10}

Thres.
& Top-$k$ & Beam & COVA
& Top-$k$ & Beam & COVA
& Top-$k$ & Beam & COVA \\
\midrule

\multicolumn{10}{c}{\textbf{Redacted}} \\
\midrule

Rec@5
&  1.54 & 1.67 & \textbf{2.88}
& 6.39 & 9.11 & \textbf{9.86}
& 0.00 & 0.30 & \textbf{0.40 }\\

Rec@25
& 6.53 & \textbf{8.84} & 8.78
& 12.53 & 17.92 & \textbf{19.83}
& 0.30 & 1.21 & \textbf{1.52} \\

Rec@50
& 11.15 & 15.50 & \textbf{17.49}
& 12.58 & 17.97 & \textbf{20.03}
& 0.30 & 1.92 & \textbf{3.04} \\

\midrule
\multicolumn{10}{c}{\textbf{Summary}} \\
\midrule

Rec@5
& 1.21 & 0.40 & \textbf{1.92}
& 5.26 & 7.79 & \textbf{8.50}
& 0.10 & 0.10 & 0.10 \\

Rec@25
& 3.14 & 4.76 & \textbf{10.43}
& 8.40 & 15.28 & \textbf{19.84}
& 0.40 & 0.20 & \textbf{0.61} \\

Rec@50
& 7.79 & 8.60 & \textbf{18.12}
& 8.50 & 15.28 & \textbf{20.95}
& 0.51 & 0.71 & \textbf{0.91} \\

\midrule
\multicolumn{10}{c}{\textbf{Name}} \\
\midrule

Rec@5
& 1.02 & 0.15 & \textbf{1.90}
& 5.43 & 4.47 & \textbf{8.49}
& 0.00 & 0.00 & 0.00 \\

Rec@25
& 2.92 & 4.81 & \textbf{9.91}
& 9.19 & 12.26 & \textbf{18.21}
& 0.00 & 0.00 & 0.00 \\

Rec@50
& 7.73 & 5.98 & \textbf{12.97}
& 9.54 & 12.26 & \textbf{19.79}
& 0.00 & 0.00 & \textbf{0.10} \\

\bottomrule
\end{tabular}
\end{table}

\begin{table*}[t]
\centering
\footnotesize
\setlength{\tabcolsep}{1.8pt}
\caption{Reconstruction performance (\%) on Qwen-2.5-7B-Instruct for city, email, and DOB across
attacker-knowledge settings on the medical dataset. $\Delta$ denotes the
percentage-point increase in PII reconstruction after fine-tuning
(FT $-$ Base). Fine-tuning increases PII reconstruction over the base model
in nearly every setting.}
\label{tab:pii_results_medical}

\begin{tabular}{ll*{6}{ccc}}
\toprule
& &
\multicolumn{3}{c}{Redacted}
& \multicolumn{3}{c}{In-distribution}
& \multicolumn{3}{c}{Summary}
& \multicolumn{3}{c}{Keywords}
& \multicolumn{3}{c}{Demographics}
& \multicolumn{3}{c}{Name} \\
\cmidrule(lr){3-5}
\cmidrule(lr){6-8}
\cmidrule(lr){9-11}
\cmidrule(lr){12-14}
\cmidrule(lr){15-17}
\cmidrule(lr){18-20}

PII & Threshold
& Base & FT & $\Delta$
& Base & FT & $\Delta$
& Base & FT & $\Delta$
& Base & FT & $\Delta$
& Base & FT & $\Delta$
& Base & FT & $\Delta$ \\
\midrule

\multirow{3}{*}{City}
& Recall@5
& 0.78 & 2.53 & \textbf{1.75}
& 1.46 & 2.77 & \textbf{1.31}
& 0.97 & 2.25 & \textbf{1.28}
& 1.32 & 2.51 & \textbf{1.19}
& 1.31 & 2.92 & \textbf{1.60}
& 0.44 & 2.48 & \textbf{2.04} \\

& Recall@25
& 3.81 & 11.73 & \textbf{7.92}
& 7.43 & 11.81 & \textbf{4.37}
& 6.80 & 10.40 & \textbf{3.60}
& 7.39 & 11.25 & \textbf{3.86}
& 7.87 & 9.33 & \textbf{1.46}
& 1.16 & 9.18 & \textbf{8.02} \\

& Recall@50
& 5.37 & 21.50 & \textbf{16.13}
& 9.18 & 22.01 & \textbf{12.83}
& 8.65 & 20.17 & \textbf{11.52}
& 10.42 & 21.16 & \textbf{10.74}
& 10.50 & 18.95 & \textbf{8.45}
& 1.46 & 14.14 & \textbf{12.68} \\

\cmidrule(lr){1-20}

\multirow{3}{*}{Email}
& Recall@5
& 0.45 & 5.77 & \textbf{5.32}
& 0.79 & 6.39 & \textbf{5.60}
& 0.83 & 4.18 & \textbf{3.36}
& 1.75 & 4.54 & \textbf{2.79}
& 2.28 & 4.64 & \textbf{2.36}
& 0.79 & 3.68 & \textbf{2.89} \\

& Recall@25
& 3.79 & 18.21 & \textbf{14.43}
& 3.77 & 18.74 & \textbf{14.97}
& 3.04 & 16.96 & \textbf{13.92}
& 3.45 & 18.70 & \textbf{15.25}
& 5.17 & 18.48 & \textbf{13.31}
& 3.06 & 11.38 & \textbf{8.32} \\

& Recall@50
& 5.20 & 20.33 & \textbf{15.13}
& 4.82 & 20.67 & \textbf{15.85}
& 4.94 & 19.68 & \textbf{14.74}
& 5.11 & 20.95 & \textbf{15.84}
& 9.28 & 20.93 & \textbf{11.65}
& 3.85 & 12.78 & \textbf{8.93} \\

\cmidrule(lr){1-20}

\multirow{3}{*}{DOB}
& Recall@5
& 0.00 & 0.48 & \textbf{0.48}
& 0.00 & 0.00 & \textbf{0.00}
& 0.00 & 0.44 & \textbf{0.44}
& 0.00 & 0.29 & \textbf{0.29}
& 0.00 & 0.10 & \textbf{0.10}
& 0.00 & 0.19 & \textbf{0.19} \\

& Recall@25
& 0.00 & 1.49 & \textbf{1.49}
& 0.00 & 0.10 & \textbf{0.10}
& 0.07 & 0.58 & \textbf{0.51}
& 0.00 & 0.44 & \textbf{0.44}
& 0.00 & 0.49 & \textbf{0.49}
& 0.10 & 0.29 & \textbf{0.19} \\

& Recall@50
& 0.00 & 2.68 & \textbf{2.68}
& 0.10 & 0.10 & \textbf{0.00}
& 0.07 & 1.07 & \textbf{1.00}
& 0.00 & 0.66 & \textbf{0.66}
& 0.00 & 0.78 & \textbf{0.78}
& 0.19 & 0.29 & \textbf{0.10} \\

\bottomrule
\end{tabular}
\end{table*}

\noindent\textbf{Prompt-Side Baselines.}
\label{appendix_prompt_side}
Table~\ref{tab:fewshot_comparison_legal} compares reconstruction with and without public few-shot demonstrations across three decoding methods on the legal dataset. The results are consistent with Section~\ref{eval:prompt_side_baseline}: public few-shot demonstrations do not reliably improve reconstruction of PII associations learned from proprietary SFT data, particularly for beam search and COVA. Top-$k$ sampling shows minor improvements.

\noindent \textbf{Decoding Baselines.}
\label{a_decoding}
To evaluate COVA on a different model, we repeat the decoding-baseline comparison using Llama~3.1~8B on the medical dataset under the same attacker-knowledge settings. The results are consistent with Section~\ref{eval:decoding_baseline}: COVA outperforms the strongest baseline across nearly all attacker-knowledge settings, and its advantage is generally more pronounced under weaker attacker knowledge. For example, COVA improves email reconstruction from 12.26\% to 19.79\% in the name-only setting at Recall@50, compared with 17.97\% to 20.03\% under the redacted setting.

\subsection{Targeted PII Reconstruction on Aligned Models.}
\label{a_instruct}
We perform the same attacks on instruction-following models to examine the effect of prior alignment on reconstruction success. Table~\ref{tab:pii_results_medical} reports reconstruction results across all PII types for the medical Q\&A at three thresholds, Recall@5, Recall@25, and Recall@50, under the attacker knowledge settings in Table~\ref{tab:attacker_knowledge}. 

The results follow a similar pattern to those for models fine-tuned from pre-trained models in Section~\ref{attack_result}. Fine-tuning increases PII reconstruction in all settings, demonstrating SFT-induced memorization beyond the model's prior or generic PII predictability. 

Partial attacker knowledge generally improves the fine-tuned model's absolute reconstruction rate relative to the name-only setting, particularly at Recall@25 and Recall@50. For email reconstruction, partial knowledge such as symptom summaries and keywords yields performance comparable to the redacted setting and consistently exceeds the name-only setting, which represents the weakest attacker knowledge.

City reconstruction requires a more careful interpretation. Although the fine-tuned model achieves its lowest city reconstruction under name-only
knowledge at Recall@50, the corresponding FT--Base increase is comparatively large. This appears to result from the unusually low base-model reconstruction rate in the name-only setting, where candidate generation sometimes collapses and produces too few valid city candidates for evaluation. The large FT--Base gap therefore reflects base-model performance suppressed by candidate-generation collapse, rather than stronger reconstruction under weaker attacker knowledge.

\subsection{PII Scrubbing Accuracy.}
\label{app:scrubbing_accuracy}
Table~\ref{tab:scrubbing_accuracy} evaluates the performance of Microsoft Presidio in detecting and masking PII. We evaluate two matching criteria. Under \emph{Exact Type}, a detection is correct only if the PII is masked and assigned the correct entity type. Under \emph{Any Type}, a detection is correct whenever the PII is masked, regardless of the assigned type.

For either criterion, precision is the fraction of predicted PII spans that correspond to true PII, while recall is the fraction of ground-truth PII spans that are successfully detected and masked:
\[
\mathrm{Precision}=\frac{\mathrm{TP}}{\mathrm{TP}+\mathrm{FP}},
\qquad
\mathrm{Recall}=\frac{\mathrm{TP}}{\mathrm{TP}+\mathrm{FN}}.
\]
where TP, FP, and FN denote true positives, false positives, and false
negatives, respectively. \emph{Exact-Type} performance therefore captures both PII detection and correct type assignment, whereas \emph{Any-Type} performance measures PII-removal coverage while ignoring type-label errors.

\begin{table}[htbp]
\centering
\footnotesize
\setlength{\tabcolsep}{3pt}
\caption{PII scrubbing performance for the medical and legal datasets.
Prec. and Rec. denote precision and recall, respectively. Exact Type requires the correct PII type, whereas Any Type ignores the predicted type.}
\label{tab:scrubbing_accuracy}

\begin{tabular}{l*{4}{cc}}
\toprule
&
\multicolumn{4}{c}{\textbf{Medical}} &
\multicolumn{4}{c}{\textbf{Legal}} \\
\cmidrule(lr){2-5}
\cmidrule(lr){6-9}

&
\multicolumn{2}{c}{\textbf{Exact Type}} &
\multicolumn{2}{c}{\textbf{Any Type}} &
\multicolumn{2}{c}{\textbf{Exact Type}} &
\multicolumn{2}{c}{\textbf{Any Type}} \\
\cmidrule(lr){2-3}
\cmidrule(lr){4-5}
\cmidrule(lr){6-7}
\cmidrule(lr){8-9}

\textbf{PII Type}
& \textbf{Prec.} & \textbf{Rec.}
& \textbf{Prec.} & \textbf{Rec.}
& \textbf{Prec.} & \textbf{Rec.}
& \textbf{Prec.} & \textbf{Rec.} \\
\midrule

Person
& 0.69 & 0.99
& 0.69 & 1.00
& 0.69 & 0.99
& 0.69 & 1.00 \\

Email
& 1.00 & 1.00
& 1.00 & 1.00
& 1.00 & 1.00
& 1.00 & 1.00 \\

Phone
& 0.98 & 0.78
& 0.99 & 0.99
& 0.98 & 0.78
& 0.99 & 0.99 \\

URL
& 0.63 & 1.00
& 0.63 & 1.00
& 0.63 & 1.00
& 0.64 & 1.00 \\

Location
& 0.52 & 0.53
& 0.56 & 0.62
& 0.52 & 0.53
& 0.56 & 0.62 \\

Date\_Time
& 0.13 & 1.00
& 0.13 & 1.00
& 0.13 & 1.00
& 0.13 & 1.00 \\

\bottomrule
\end{tabular}
\end{table}

\noindent \textbf{PII Types.}
Scrubbing quality varies substantially across PII types. Identifiers with more regular formats are generally detected more reliably: email achieves perfect precision and recall across both datasets, while phone reaches near-perfect precision and recall under the \emph{Any-Type} criterion. In contrast, location exhibits substantially weaker performance, with precision and recall between 0.52 and 0.62. Location generally varies in different formats, such as cities, street names, or full addresses.

\noindent \textbf{Overmasking.} Some PII types achieve nearly perfect recall at the cost of low precision. \texttt{DATE\_TIME} is the clearest example, with recall of 1.00 but precision of only 0.13 across both datasets, indicating extensive overmasking. Qualitative inspection helps explain this pattern: the scrubber frequently labels unrelated numeric expressions, including medication quantities and ZIP codes, as \texttt{DATE\_TIME}. Person and URL also exhibit lower precision than email, indicating that many non-PII spans are incorrectly identified and masked as PII.

Overall scrubbing quality varies substantially across PII types: structured PII such as email are detected reliably whereas locations are often missed and \texttt{DATE\_TIME} produces high false positives. These results demonstrate that scrubbing can leave some PII unmasked and incorrectly remove non-PII content, affecting the downstream model utility.

\end{document}